\documentclass[lettersize,journal]{IEEEtran}
\usepackage{amsmath,amsfonts}
\usepackage{algorithmic}
\usepackage{algorithm}

\usepackage{amssymb}
\usepackage{array}
\usepackage[caption=false,font=footnotesize,labelfont=rm,textfont=rm]{subfig}
\usepackage{textcomp}
\usepackage{stfloats}
\usepackage{url}
\usepackage{verbatim}
\usepackage{graphicx}
\usepackage{cite}

\usepackage[hidelinks]{hyperref}
\usepackage{amsthm}

\theoremstyle{definition}

\IEEEaftertitletext{\vspace{-20pt}}
\renewcommand{\appendixname}{Appendix}
\begin{document}

\title{Robust Beamforming for Multiuser MIMO Systems with Unknown Channel Statistics: A Hybrid Offline-Online Framework}

\author{\IEEEauthorblockN{Wenzhuo Zou, \IEEEmembership{Student Member, IEEE}, Ming-Min Zhao, \IEEEmembership{Senior Member, IEEE}, An Liu, \IEEEmembership{Senior Member, IEEE}, and Min-Jian Zhao, \IEEEmembership{Member, IEEE} \thanks{W. Zou, M. M. Zhao, A. Liu, and M. J. Zhao are with the College of Information Science and Electronic Engineering, Zhejiang University, Hangzhou 310027, China, and also with Zhejiang Provincial Key Laboratory of Multi-Modal Communication Networks and Intelligent Information Processing, Hangzhou 310027, China (e-mail: \{12231090, zmmblack, anliu, mjzhao\}@zju.edu.cn).}
	}
}
\markboth{}%
{}

\IEEEpubid{}

\maketitle

\begin{abstract}
Robust beamforming design under imperfect channel state information (CSI) is a fundamental challenge in multiuser multiple-input multiple-output (MU-MIMO) systems, particularly when the channel estimation error statistics are unknown. Conventional model-driven methods usually rely on prior knowledge of the error covariance matrix and data-driven deep learning approaches suffer from poor generalization capability to unseen channel conditions. To address these limitations, this paper proposes a hybrid offline-online framework that achieves effective offline learning and rapid online adaptation.
In the offline phase, we propose a shared (among users) deep neural network (DNN) that is able to learn the channel estimation error covariance from observed samples, thus enabling robust beamforming without statistical priors. Meanwhile, to facilitate real-time deployment, we propose a sparse augmented low-rank (SALR) method to reduce complexity while maintaining comparable performance.
In the online phase, we show that the proposed network can be rapidly fine-tuned with minimal gradient steps.
Furthermore, a multiple basis model-agnostic meta-learning (MB-MAML) strategy is further proposed to maintain multiple meta-initializations and by dynamically selecting the best one online, we can improve the adaptation and generalization capability of the proposed framework under unseen or non-stationary channels.
Simulation results demonstrate that the proposed offline-online framework exhibits strong robustness across diverse channel conditions and it is able to significantly outperform state-of-the-art (SOTA) baselines.
\end{abstract}

\begin{IEEEkeywords}
Robust beamforming, CSI, MIMO, deep learning, meta-learning.
\end{IEEEkeywords}

\section{INTRODUCTION}
\label{sec:introduction}
\IEEEPARstart{M}{ultiuser} multiple-input-multiple-output (MU-MIMO) technology has fundamentally reshaped the landscape of modern wireless communications by enabling spatial multiplexing, interference suppression, and enhanced spectral efficiency \cite{6736761, 7064850}. At the heart of MIMO transceiver design lies beamforming, a technique that tailors transmit signals to exploit channel structure, thereby maximizing system utility, such as weighted sum rate (WSR), under practical constraints like total or per-antenna power limits. Classical beamforming typically assumes perfect channel state information (CSI) at the transmitter; however, in real-world deployments, CSI is inevitably imperfect due to estimation noise, quantization, feedback delay, and user mobility, etc. This mismatch motivates the development of advanced beamforming strategies that can maintain performance under channel uncertainty, a challenge further amplified in dense, interference-limited multiuser scenarios.

The design of transmit beamforming in multiuser MIMO systems has evolved through distinct phases, each addressing limitations of the prior. It begins with maximum ratio transmission (MRT) \cite{6812124MRT}, which focuses solely on boosting signal power toward intended users and is effective in noise-limited scenarios but ignoring interference entirely. To counter inter-user interference, zero-forcing (ZF) beamforming was introduced \cite{6093291ZF}, forcing interference to zero via channel inversion; however, this often amplifies noise, especially in poorly conditioned channels. Recognizing this trade-off, regularized ZF (RZF) emerged \cite{8668481RZF}, which adds a regularization term to balance signal gain, interference suppression, and noise resilience. Numerous other beamforming strategies have also been proposed in previous works, e.g., \cite{6928432, 8118138, 7104128}. While these linear methods offer low complexity, they do not directly optimize system-level objectives like WSR. For that purpose, optimization-based frameworks are required, e.g., successive convex approximation (SCA) \cite{5205801, 5752426SCA}, which iteratively solves non-convex problems through local convex surrogates. Among its most successful realizations is the weighted minimum mean squared error (WMMSE) algorithm \cite{shi2011iteratively}, which embeds physical receiver structure into the approximation, yielding efficient updates and near-optimal performance. A wide range of optimization-driven robust beamforming approaches have also been developed in \cite{8690820, 8756668}. While effective, these methods often involve iterative optimization and high computational overhead, which paves the way for learning-based alternatives that bypass explicit algorithmic design.
\IEEEpubidadjcol

Recent works have embraced deep learning to replace iterative optimization with direct end-to-end mapping from CSI to beamforming vectors. For instance, the work \cite{sunLearningOptimizeTraining2018b} applied the multi-layer perceptron (MLP) to approximate the iterative WMMSE algorithm in a multiuser single-input single-output system. The work \cite{9322310} introduced a method that uses the WMMSE algorithm to generate labels for offline supervised learning, combined with online unsupervised learning using the WSR as the loss function.
However, data-driven approaches present notable limitations, e.g., vulnerability to output inaccuracies caused by data noise or model bias, and inherent difficulties in guaranteeing robustness and generalization capabilities. To simultaneously address beamforming performance, robustness, and computational efficiency, researchers have turned to domain expertise-infused model-driven neural network architectures for beamforming design. Specifically, the work \cite{xiaDeepLearningFramework2020} proposed a model-driven deep learning framework that integrates expert knowledge of the WMMSE algorithm into the network architecture, significantly reducing the number of trainable parameters and enhancing interpretability. The work \cite{huIterativeAlgorithmInduced2021a} introduced an iterative algorithm-induced deep learning approach that unfolds the iterations of the WMMSE algorithm into a layer-wise structure, replacing complex matrix operations with lightweight neural networks to reduce computational complexity while maintaining performance. 
Other deep learning-based beamforming approaches have also been explored in the literature, e.g., \cite{attiahDeepLearningChannel2022a, 9414561,9337188, yuanTransferLearningMeta2021}.

While the aforementioned algorithms offer attractive convergence properties, their performance critically hinges on the availability of perfect CSI, which is an assumption rarely met in practice. In real-world deployments, CSI is often imperfect due to channel estimation errors, pilot contamination, and hardware impairments, making robust beamforming design essential to mitigate performance degradation. Early approaches enhance linear beamformers under uncertainty, e.g., the work \cite{6155708} proposed a robust RZF scheme that optimizes the regularization factor to maximize the average signal-to-interference-plus-noise-ratio (SINR) in the presence of channel estimation errors. For better performance, optimization-theory-driven methods are employed, for example, the work \cite{razaviyaynStochasticWeightedMMSE2013} introduced the stochastic weighted minimum mean square error (SWMMSE) algorithm, which iteratively optimizes the average WSR under statistical CSI uncertainty, but it typically requires a substantial number of CSI samples and multiple iterations, introducing significant computational overhead. To address this, the work \cite{8694866} derived deterministic equivalents of the average WSR and employs the majorization-minimization (MM) method for efficient optimization. Further complexity reduction is achieved in \cite{wangRobustWMMSEBasedPrecoder2024}, which exploited channel sparsity in the millimeter-wave band to simplify the deterministic equivalent computation. Yet these methods still rely on various statistical or structural priors and incur high computational cost.

In the context of robust beamforming, deep learning has emerged as a powerful tool to embed algorithmic priors or uncertainty modeling directly into neural architectures, circumventing the computational burden of iterative optimization while enhancing generalization under imperfect CSI. The work \cite{9369390} proposed a deep neural network (DNN)-based approach that jointly leverages imperfect CSI and its statistical properties to design robust beamformers, effectively encoding distributional assumptions into the training process. The work \cite{shiRobustWMMSEPrecoder2023} developed a structure-aware learning framework that captures the intrinsic low-dimensional geometry of robust beamforming vectors, learning compact parameterizations directly from available CSI through a neural network, thus reducing model complexity while preserving performance. Recently, the work \cite{wangRobustWMMSEBasedPrecoder2024} unfolded a robust beamforming algorithm into a layer-wise neural network, introduced trainable matrices to compensate for approximation loss and accelerate convergence. While these methods significantly outperform black-box neural beamformers in uncertain environments, they remain fundamentally reliant on prior knowledge, whether statistical (e.g., error distribution in \cite{9369390}), structural (e.g., low-dimensional manifold in \cite{shiRobustWMMSEPrecoder2023}), or algorithmic (e.g., unfolding structure in \cite{wangRobustWMMSEBasedPrecoder2024}). This implicit dependency limits their robustness in open-set or non-stationary deployments and this motivates us to propose a prior-free neural beamforming framework based on meta-learning, which requires no statistical, structural, or algorithmic prior knowledge. Other notable learning-based approaches to robust beamforming have also been explored in the literature \cite{10879353, 10739925, 11143219}.

To overcome the above drawbacks of existing methods, we propose a hybrid offline-online robust beamforming framework that operates without any prior statistical knowledge of channel estimation errors. The core innovation is a two-phase learning paradigm: an offline meta-training phase that learns a generalizable representation of channel error statistics from diverse scenarios, followed by an online adaptation phase that rapidly fine-tunes this representation using only a few gradient steps on newly observed channel samples. This synergistic design enables real-time, robust performance under unknown and non-stationary channel conditions.
The main contributions of this paper are summarized as follows: 
\begin{itemize}
\item We propose a hybrid offline-online framework for robust beamforming that requires no prior knowledge of channel error statistics. This two-phase design enables statistically agnostic robustness by learning statistical priors offline and rapidly adapting online with minimal gradient steps.

\item We propose a sparse augmented low-rank (SALR) method to decompose the predicted covariance matrix into low-rank and sparse components. This method reduces model complexity by orders of magnitude while preserving estimation accuracy, thus enabling real-time, low-overhead online deployment.

\item We develop a multiple basis model-agnostic meta-learning (MB-MAML) strategy, which maintains a diverse set of meta-initializations during offline training and dynamically selects or interpolates the most suitable one during online adaptation. This significantly improves generalization under distribution shift and non-stationary channel conditions.

\item Extensive simulations demonstrate that the proposed framework achieves great WSR performance under imperfect CSI, consistently outperforming state-of-the-art robust and data-driven baselines across practical signal-to-noise ratio (SNR) and error regimes.
\end{itemize}

The remainder of this paper is organized as follows. Section~\ref{system model and problem formulations} presents the system model and formulates the robust beamforming design as an average WSR maximization problem under imperfect CSI. Section~\ref{sec:related work} briefly reviews conventional robust beamforming approaches. Section~\ref{sec:offline_robust_beamforming_design} introduces the proposed hybrid offline-online robust beamforming framework. Section~\ref{sec:simulation_results} provides extensive numerical evaluations and conclusions are drawn in Section~\ref{sec:conclusion}.

\emph{Notations}: this  paper uses lower case, lower case boldface, and upper case boldface letters to denote scalars,
vectors, and matrices, respectively. 
$\mathbb{C}^{M \times N}(\mathbb{R}^{M \times N})$ denotes the $M \times N$ dimensional complex (real) matrix space, $(\cdot)^H$, $(\cdot)^T$, and $(\cdot)^*$ denote conjugate transpose, transpose, and complex conjugate operations, respectively.
$\text{vec}(\mathbf{A})$ denotes the vectorization of matrix $\mathbf{A}$ by stacking its columns into a single column vector, $\otimes$ denotes the Kronecker product, $\mathbb{E}\{\cdot\}$ denotes the statistical expectation operation, $\operatorname{Tr}(\cdot)$ denotes the trace of a matrix, $\|\cdot\|_F$ denotes the Frobenius norm of a matrix, $\mathbf{I}_M$ denotes the $M \times M$ identity matrix, and $\mathcal{CN}(\mathbf{m},\mathbf{R})$ denotes the circularly symmetric complex Gaussian distribution with mean vector $\mathbf{m}$ and covariance matrix $\mathbf{R}$.

\section{SYSTEM MODEL AND PROBLEM FORMULATION}
\label{system model and problem formulations}
\subsection{System Model}
\label{system model}
\begin{figure}
\centering
\includegraphics[width=0.8\linewidth]{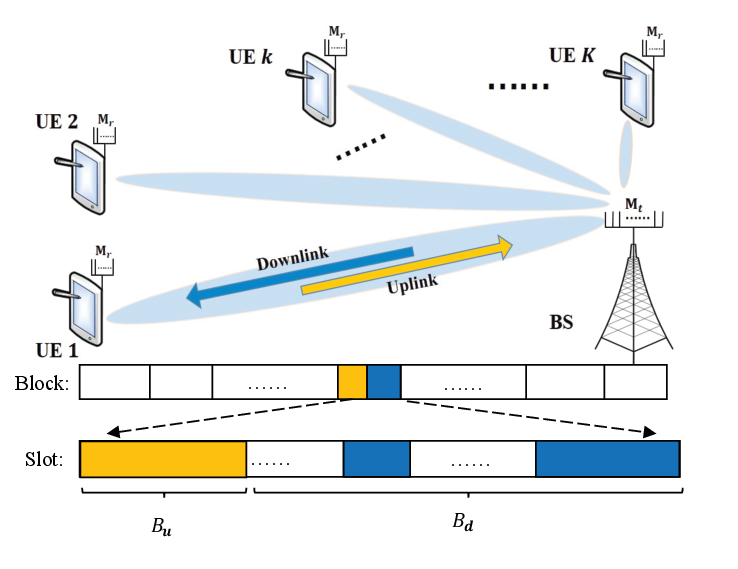} \vspace{-1em}
\caption{Massive MU-MIMO system model.}
\label{fig1 system_model}
\end{figure}

We consider a MU-MIMO downlink system, as illustrated in Fig.~\ref{fig1 system_model}, where a base station (BS) equipped with an $M_t$-antenna uniform linear array (ULA) serves $K$ single-antenna users. The system operates in time-division duplexing (TDD) mode, with each coherence block divided into $B_u + B_d$ slots. Specifically, the first $B_u$ slots are allocated for uplink training, and the remaining $B_d$ slots for downlink data transmission.
Let $\mathbf{h}_{k,b} \in \mathbb{C}^{M_t \times 1}$ denote the flat-fading channel vector between the BS and user $k$ during the $b$-th slot. Assuming channel coherence over each block, the slot index $b$ in $\mathbf{h}_{k,b}$ is omitted for notational simplicity, yielding $\mathbf{h}_{k}$. Besides, by further assuming channel reciprocity, the BS is able to acquire the  downlink CSI via uplink pilot transmission \cite{hoydisMassiveMIMOUL2013}.

Let $\mathbf{Y}_b^{\text{BS}} \in \mathbb{C}^{M_t \times L}$ denote the received signal matrix at the BS during the $b$-th training slot, given by
\begin{equation}
  \label{eq:received_matrix}
  \mathbf{Y}_b^{\text{BS}} = \sum_{k=1}^{K} \mathbf{h}_{k} \mathbf{x}_{kb}^H + \mathbf{Z}_b,
\end{equation}
where $\mathbf{x}_{kb} \in \mathbb{C}^{L \times 1}$ denotes the pilot symbols of user $k$ with pilot length $L$, $\mathbf{Z}_b$ represents a noise random matrix with each element following an independent and identically distributed (i.i.d.) complex Gaussian distribution with zero mean and variance $\sigma_{\text{BS}}^2$.
By vectorizing the received matrix $\mathbf{Y}_b^{\text{BS}}$ column-wise,
we obtain
\begin{equation}
  \label{eq:vec_received_matrix}
  \text{vec}(\mathbf{Y}_b^{\text{BS}}) = \sum_{k=1}^{K} (\mathbf{x}^{*}_{kb} \otimes \mathbf{I}_{M_t})  \mathbf{h}_{k} + \text{vec}( \mathbf{Z}_b ).
\end{equation}
Assuming that orthogonal pilots are employed, we have $\mathbf{x}_{kb}\mathbf{x}^{H}_{kb}=\mathbf{I}_{L}$ and $\mathbf{x}_{kb}\mathbf{x}^{H}_{lb}=\mathbf{0}$ for $l\neq k$. This enables separation of the received signal matrix for each user, i.e.,
\begin{equation}
  \label{received signal matrix for each user}
  ({\mathbf{x}^{T}_{kb}} \otimes \mathbf{I}_{M_t}) \text{vec}(\mathbf{Y}_b^{\text{BS}}) 
= \mathbf{h}_{k} + ({\mathbf{x}^{T}_{kb}} \otimes \mathbf{I}_{M_t}) \text{vec}( \mathbf{Z}_b ).
\end{equation}
The channel can be estimated using either the least squares (LS) or the linear minimum mean square error (LMMSE) criterion as follows:
\begin{align}
\hat{\mathbf{h}}^{\text{LS}}_{k} &= \left( \mathbf{x}_{kb}^T \otimes \mathbf{I}_{M_t} \right)\mathrm{vec}\left(\mathbf{Y}_b^{\text{BS}}\right),
\end{align}
\begin{align}
\hat{\mathbf{h}}^{\text{LMMSE}}_{k} &= \mathbf{R}_{\mathbf{h}_k}\left(\mathbf{R}_{\mathbf{h}_k}+L \sigma^2_{\text{BS}}\mathbf{I}\right)^{-1}
\left(\mathbf{x}_{kb}^T \otimes \mathbf{I}_{M_t}\right)\mathrm{vec}\left(\mathbf{Y}_b^{\text{BS}}\right).
\end{align}
where $\mathbf{R}_{\mathbf{h}_k}=\mathbb{E}\{\mathbf{h}_k\mathbf{h}_k^H\}$ is the channel covariance matrix of user $k$.
The LS and LMMSE are just used as examples of channel estimation methods in this paper, and other channel estimation methods such as compressed sensing based methods and learning-based methods can also be applied in our proposed framework.
Regardless of the methods used for estimation, the channel estimation error is defined as the difference between the true channel and its estimate, i.e.,
\begin{equation}
  \label{eq:estimation_error}
  \mathbf{e}_k = \mathbf{h}_k - \hat{\mathbf{h}}_k.
\end{equation}
In many practical robust design frameworks, the estimation error $\mathbf{e}_k$ is modeled as a zero-mean complex Gaussian random vector, i.e., $\mathbf{e}_k \sim \mathcal{CN}(\mathbf{0}, \mathbf{R}_{\mathbf{e}_k})$, where the covariance matrix $\mathbf{R}_{\mathbf{e}_k}$ captures the aggregate uncertainty from both estimation noise and any residual errors. This Gaussian model, while analytically tractable, relies on accurate knowledge of system parameters and channel statistics.
However, in real-world deployments, $\mathbf{R}_{\mathbf{e}_k}$ is influenced by multiple practical factors, including the choice of estimator (e.g., LS or LMMSE), pilot length $L$, pilot power, uplink SNR, user density (which induces pilot contamination), imperfections in practical hardware chains, and channel dynamics such as Doppler spread in high-mobility scenarios. Since it is rarely known a priori, this uncertainty in the error statistics motivates our hybrid offline-online framework.
Specifically, in challenging conditions such as low SNR or high mobility, where single-shot estimates are highly unreliable, we perform $N$ independent channel estimations (in our simulations, we set $N=2$) to collect a set of channel samples $\{ \hat{\mathbf{h}}_{k,n} \}_{n=1}^N$ where $\hat{\mathbf{h}}_{k,n}$ represents the channel estimate at the $n$-th trial. These samples serve as inputs to our hybrid offline-online architecture, enabling data-driven inference of the error covariance $\mathbf{R}_{\mathbf{e}_k}$ without requiring prior knowledge of channel statistics.

We now consider the downlink data transmission phase within the same coherence block.
The BS transmits independent data symbol $s_k \in \mathbb{C}$ to each user $k$ using the beamforming vector $\mathbf{v}_k \in \mathbb{C}^{M_t \times 1}$, where $\mathbb{E}\{|s_k|^2\} = 1$. The received signal at user $k$ is given by
\begin{equation}
  \label{eq:received_signal}
  y_k = \mathbf{h}_k^H \mathbf{v}_k s_k + \sum_{i \neq k} \mathbf{h}_k^H \mathbf{v}_i s_i + n_k,
\end{equation}
where $n_k \sim \mathcal{CN}(0, \sigma_k^2)$ denotes the additive white Gaussian noise at the user $k$.
Based on \eqref{eq:received_signal}, the achievable average rate for user $k$ is expressed as
\begin{equation}
  \label{eq:rate}
  \mathbb{E} \{\mathcal{R}_{k} \} =\mathbb{E} \left\{ \log_2 \left(1 + \frac{\left|\mathbf{h}_{k}^{H}\mathbf{v}_{k}\right|^{2}}{\sum_{i \neq k}^{K}\left|\mathbf{h}_{k}^{H}\mathbf{v}_{i}\right|^{2} + \sigma_{k}^{2}}\right) \right\}.
\end{equation}
where the expectation is taken over the distribution of the channels $\{ \mathbf{h}_k \}_{k=1}^K$.

\subsection{Problem Formulation}
In this work, we focus on robust beamforming design and the following average WSR maximization problem is formulated subject to a total transmit power constraint:
\begin{equation*}
\label{eq:robust_beamforming_problem}
\text{P1:}\quad\max_{\{\mathbf{v}_{k}\}}\,\sum_{k=1}^{K}\omega_{k}\mathbb{E}\left\{\mathcal{R}_{k}\right\},
\quad\text{s.t.}\quad\sum_{k=1}^{K}\operatorname{Tr}(\mathbf{v}_{k}\mathbf{v}_{k}^{H}) \leq P_{\max}
\end{equation*}
where $\omega_k > 0$ denotes the priority weight of user $k$, and $P_{\max}$ is the maximum transmit power at the BS. Note that if the channel statistics (e.g., the prior distributions of the channel and noise) are perfectly known, the relationship between the observed pilots and the channel $\mathbf{h}_k$ can be characterized analytically, yielding estimators such as the LMMSE. However, in this work, such statistics are considered unknown, and we therefore propose a data-driven, distribution-agnostic approach.
Moreover to facilitate gradient-based optimization in our framework, we further reformulate \text{P1} into an equivalent unconstrained problem by incorporating the power constraint into the objective via a penalty term \cite{huIterativeAlgorithmInduced2021a}:\footnote{Note that the optimal solution $\mathbf{v}_k^*$ of original problem \text{P1} and the optimal solution $\mathbf{v}_k^{\star}$ of the transformed problem \text{P2} satisfy the following relation: $\mathbf{v}_k^* = \alpha \mathbf{v}_k^{\star}$, where $\alpha = \sqrt{\frac{P_{\max}}{\sum_{k=1}^{K}\operatorname{Tr}(\mathbf{v}_{k}^{\star} (\mathbf{v}_{k}^{\star})^{H})}}$ is a scaling factor that ensures the power constraint is satisfied.}
\begin{equation*}
\label{eq:unconstrained_sum_rate_maximization}
\text{P2:}\quad\max_{\{\mathbf{v}_{k}\}}\,\sum_{k=1}^{K}\omega_{k}\mathbb{E}\left\{\mathcal{\tilde{R}}_{k}\right\}
\end{equation*}
where 
\begin{equation*}
\label{eq:tilde_rate}
\mathcal{\tilde{R}}_{k} = \log\left(1 + \frac{\left|\mathbf{h}_{k}^{H}\mathbf{v}_{k}\right|^{2} } {\sum_{i \neq k}^{K}\left|\mathbf{h}_{k}^{H}\mathbf{v}_{i}\right|^{2} + \frac{\sigma_{k}^{2}}{P_{\max}} \sum_{k=1}^{K}\operatorname{Tr}(\mathbf{v}_{k}\mathbf{v}_{k}^{H})}  \right).
\end{equation*}
Despite this reformulation, problem P2 remains non-convex
and challenging to solve optimally because the objective involves coupled fractional quadratic terms and uncertain channel errors, which motivates us to develop efficient learning-based solutions in the latter section.
%
\section{Related Work}
\label{sec:related work}
In this section, we briefly review existing robust beamforming approaches under imperfect CSI with known channel statistics, which serve as important baselines for comparison in our simulations. We first discuss how the original problem $\text{P2}$ can be transformed into an equivalent stochastic weighted minimum mean square error (SWMMSE) problem, a well-established approach when perfect channel statistics are available \cite{razaviyaynStochasticWeightedMMSE2013}. Specifically, \text{P2} can be equivalently reformulated as
\begin{equation*}
\label{eq:wmmse_problem}
\text{P3}:\min_{\{\mathbf{v}_{k}\}} \mathbb{E} \left\{ \min_{\{{u}_k,w_{k}\}}\sum_{k = 1}^{K}\omega_{k}( - \log w_k + \mathrm{Tr}(w_{k}{\widetilde{\mathbf{E}}}_{k})) \right\}
\end{equation*}
where $u_k \in \mathbb{C}$ and $w_k > 0$ denote the introduced receive coefficient and weighting factor for user $k$, respectively, and $\widetilde{\mathbf{E}}_k$ is the mean square error (MSE) matrix defined as
\begin{multline}
\label{eq:tilde_E}
{\widetilde{\mathbf{E}}}_{k} \triangleq \left( 1 - u_{k}^{H}\mathbf{h}_{k}^{H}\mathbf{v}_{k} \right)\left( 1 - u_{k}^{H}\mathbf{h}_{k}^{H}\mathbf{v}_{k} \right)^{H} \\
+ \sum_{i \neq k}^{K}{u_{k}^{H}\mathbf{h}_{k}^{H}\mathbf{v}_{i}}\mathbf{v}_{i}^{H}\mathbf{h}_{k}u_{k} +  \frac {\sum_{k = 1}^{K}{\mathrm{Tr}(\mathbf{v}_{k}\mathbf{v}_{k}^{H})}}{ P_{\max}}\sigma^{2}u_{k}^{H}u_{k}
\end{multline}
Since the objective function of problem $\text{P3}$ does not admit a closed-form expression, the SWMMSE algorithm employs the sample average approximation (SAA) method to handle it \cite{razaviyaynStochasticWeightedMMSE2013}. Specifically, at each iteration, the receive coefficients $\{u_k\}$ and weighting factors $\{w_k\}$ are updated based on instantaneous channel realizations, while the beamforming vectors $\{\mathbf{v}_k\}$ are updated using the empirical average over multiple channel samples to approximate the stochastic expectation. Let $\mathbf{h}_{k,n}$ denote the $n$-th channel sample for user $k$. Given a batch of $N$ i.i.d. samples, the update rules for $\{u_k, w_k, \mathbf{v}_k \}$ at iteration $r$ are as follows:
\begin{equation}
\label{eq:optimal_u_SWWMSE}
u_{k}^r = \left( \sum_{i = 1}^{K} \mathbf{h}_{k,r}^{H}\mathbf{v}_{i}^{r-1}(\mathbf{v}_{i}^{r-1})^{H}\mathbf{h}_{k,r}  + \sigma^{2}_k \right)^{- 1}\mathbf{h}_{k,r}^{H}\mathbf{v}_{k}^{r-1}
\end{equation}
\begin{equation}
\label{eq:optimal_w_SWWMSE}
w_{k}^r = \left( 1 - (u_{k}^r)^{H}(\mathbf{h}_{k,r})^{H}\mathbf{v}_{k}^{r-1} \right)^{- 1}
\end{equation}
\begin{equation}
\label{eq:optimal_v_SWWMSE}
\mathbf{v}_{k}^{r} = \omega_k \left( \sum_{n = 1}^{r} \left({\sum_{i = 1}^{K}\lambda_i^n \mathbf{h}_{i,n}\mathbf{h}_{i,n}^{H}}  + \mu_n \mathbf{I} \right) \right)^{-1} \sum_{n = 1}^{r}{\mathbf{h}_{k,n}w_{k}^nu_{k}^n}
\end{equation}
where $\lambda_i^r = u_{i}^rw_{i}^r(u_{i}^r)^H$, $\mu^n = \frac{\sigma^2}{P_{\max}} \sum_{i=1}^K \lambda_i^n$ and following each iteration, power normalization is applied to $\mathbf{v}_k$ to ensure the power constraint is satisfied.
Although SWMMSE is guaranteed to converge to a stationary point of problem $\text{P3}$, it typically requires hundreds of iterations and samples to reach practical convergence, resulting in prohibitive computational complexity for real-time implementation.

Besides, the work in \cite{shiRobustWMMSEPrecoder2023} proposed a robust WMMSE algorithm that compute the expectation-related operations in closed form, thereby obviating the need for iterative Monte Carlo sampling. The resulting update rules at iteration $r$ under statistical CSI are given by
\begin{equation}
\label{eq:optimal_u_RW}
u_{k}^r = \left( \sum_{i = 1}^{K} \mathbb{E} \{ \mathbf{h}_{k}^{H}\mathbf{v}_{i}^{r-1}(\mathbf{v}_{i}^{r-1})^{H}\mathbf{h}_{k} \} + \sigma^{2}_k \right)^{- 1} \mathbf{\bar{h}}_{k}^{H}\mathbf{v}_{k}^{r-1}
\end{equation}
\begin{equation}
\label{eq:optimal_w_RW}
w_{k}^{r} = \left( 1 - (u_{k}^{r})^{H}\mathbf{\bar{h}}_{k}^{H}\mathbf{v}_{k}^{r-1} \right)^{- 1}
\end{equation}
\begin{equation}
\label{eq:optimal_v_RW}
\mathbf{v}_{k}^{r} = \omega_k \left(\sum_{i = 1}^{K} \mathbb{E}\{\lambda_i^r \mathbf{h}_{i}\mathbf{h}_{i}^{H}\}  + \mu^r \mathbf{I} \right)^{-1} {\mathbf{\bar{h}}_{k} w_{k}^r u_{k}^r}
\end{equation}
where  $\lambda_i^r = u_{i}^r w_{i}^r (u_{i}^r)^H$, $\mu^r = \frac{\sigma^2}{P_{\max}} \sum_{i=1}^K \lambda_i^r$ and $\bar{\mathbf{h}}_k = \mathbb{E}\{\mathbf{h}_k\}$ denotes the channel mean. These updates are iteratively executed until convergence, yielding a suboptimal solution to the robust WMMSE problem \text{P3}. Similar to the SWWMSE algorithm, a power normalization step is applied to $\mathbf{v}_k$ to ensure the power constraint is satisfied. While computationally efficient, this method relies on accurate knowledge of the channel statistics (e.g., $\mathbf{R}_{\mathbf{h}_k}$ and $\bar{\mathbf{h}}_k$) which does not hold in many practical scenarios.

\section{Hybrid Offline-Online Robust Beamforming Framework}
\label{sec:offline_robust_beamforming_design}

In this section, we introduce the proposed hybrid offline-online framework for robust beamforming design under unknown channel statistics. We first introduce the overall framework and network design, followed by the SALR method for lightweight covariance approximation and the MB-MAML strategy for fast online adaptation.
\begin{figure*}
\centering
\includegraphics[width=0.8\textwidth]{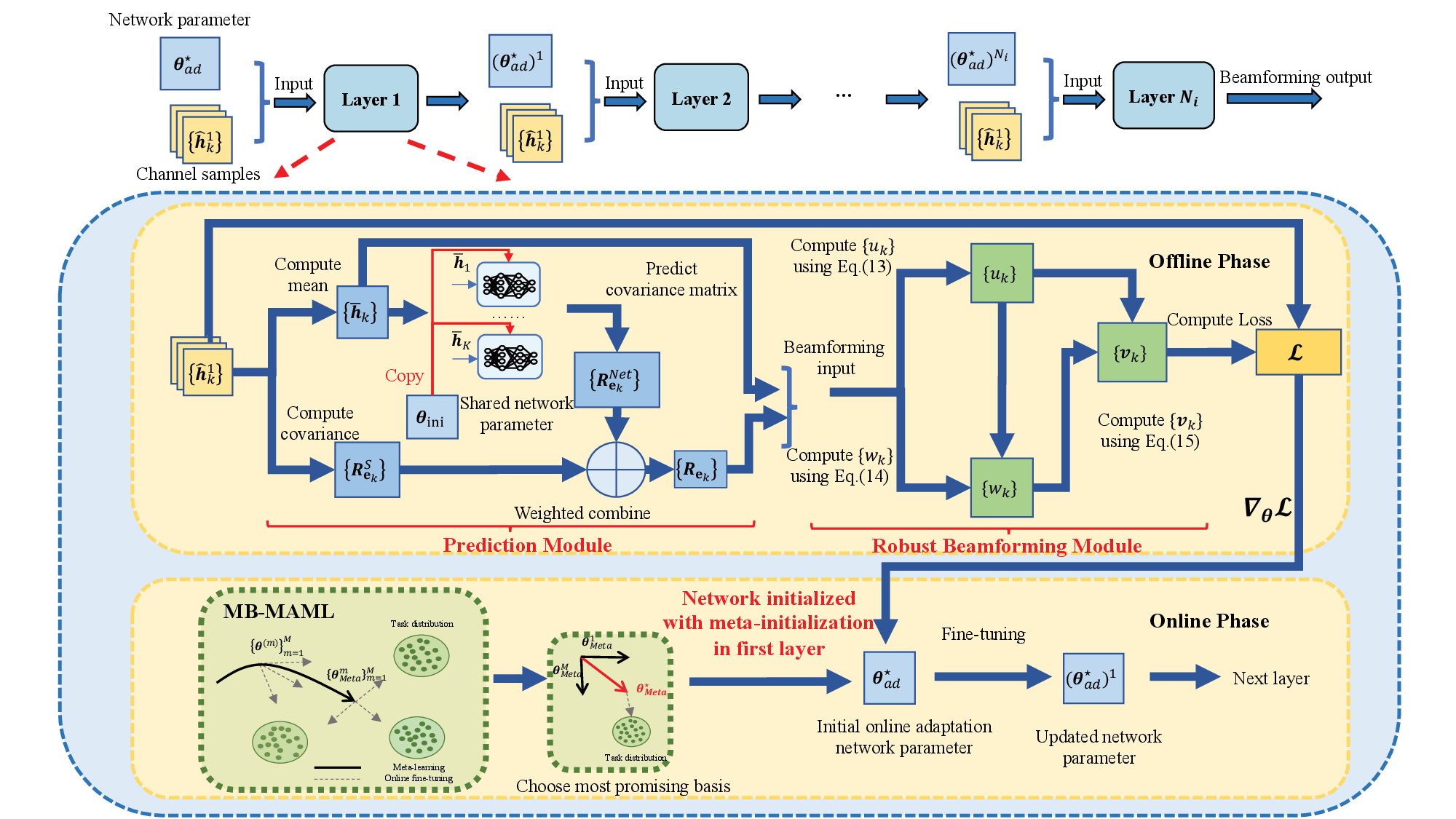}
\caption{ The complete pipeline of the proposed hybrid offline-online robust beamforming framework.}
\label{fig:total_framework_1}
\end{figure*}

\subsection{Proposed Framework}
\label{sec:proposed_framework}
As illustrated in Fig.~\ref{fig:total_framework_1}, the proposed framework consists of an offline phase and an online phase. 
In the former, the core is a robust beamforming module,
which processes generated channel samples to implicitly capture the statistical information necessary for designing beams that are resilient to channel estimation imperfections. In particular, a dedicated prediction module, employing a neural network, is integrated to provide these covariance matrices for all users directly from the channel mean estimates, which can be obtained using standard channel estimation methods, such as LS, LMMSE or more advanced compressive sensing methods \cite{8989963}.
In the latter, the prediction module is fine-tuned based on the observed channel samples, and the MB-MAML method is further leveraged to exploit multiple meta-initializations (also called meta-bases) to enable swift adaptation to varying channel conditions.

\subsubsection{Offline Learning Phase}
In this phase, the robust WMMSE method from \cite{shiRobustWMMSEPrecoder2023} is developed as the backbone, and the beamforming vectors are updated as (\ref{eq:optimal_u_RW})--(\ref{eq:optimal_v_RW}).
As can be seen, the above robust WMMSE updates depend essential on the expectations computed via 
\begin{align}
    \label{eq:second_order_expectation}
    \mathbb{E}\{\mathbf{h}_k \mathbf{h}_k^H\} &= \bar{\mathbf{h}}_k \bar{\mathbf{h}}_k^H + \mathbf{R}_{\mathbf{e}_k}, \\
    \label{eq:second_order_expectation_v}
    \mathbb{E}\{\mathbf{h}_k^H \mathbf{v}_i \mathbf{v}_i^H \mathbf{h}_k\} &= \bar{\mathbf{h}}_k^H \mathbf{v}_i \mathbf{v}_i^H \bar{\mathbf{h}}_k + \mathrm{Tr}(\mathbf{v}_i \mathbf{v}_i^H \mathbf{R}_{\mathbf{e}_k}),
\end{align}
which in turn require the channel error covariance matrices $\mathbf{R}_{\mathbf{e}_k}$. Acquiring these matrices accurately in practice is a major challenge, thus motivating the integration of a covariance prediction module to enable the proposed robust beamforming design.

We implement the covariance prediction module using a fully-connected neural network. A key feature of this network is its parameter-sharing architecture across users, which not only minimizes the model footprint but also enables the transfer of learned knowledge about error characteristics. The model is trained offline on a large training dataset $\mathcal{D}_{\text{tr}}$, which consists of a large number of channel observations, where each observation contains $N$ channel estimation samples $\{\hat{\mathbf{h}}_{k,n}\}_{n=1}^{N}$. Critically, the network's input is the empirical mean of these channel estimation samples for a given moment, which is calculated by
\begin{equation}
\label{eq:channel_mean}
\bar{\mathbf{h}}_{k} = \frac{1}{N} \sum_{n=1}^{N} \hat{\mathbf{h}}_{k,n},
\end{equation}
rather than the raw samples themselves. This yields two primary benefits: 1) it fixes the network's input size, leading to a stable and efficient computational graph during operation; 2) it forces the network to learn the underlying mapping from the central tendency of the channel to its error statistics, which enhances generalization.
The network outputs a predicted covariance matrix using network parameter $\boldsymbol{\theta}$, which is given by
\begin{equation}
    \label{eq:covariance_matrix_network}
    \mathbf{R}_{\mathbf{e}_k}^{\text{Net}} = f_{\boldsymbol{\theta}}(\bar{\mathbf{h}}_k),
\end{equation}
which is fused with the sample covariance $\mathbf{R}_{\mathbf{e}_k}^{S}$ to form the final covariance estimate, i.e.,
\begin{equation}
    \label{eq:covariance_matrix_samples}
  \mathbf{R}_{\mathbf{e}_k}^{S} = \frac{1}{N} \sum_{n=1}^{N} (\hat{\mathbf{h}}_{k,n} - \bar{\mathbf{h}}_k)(\hat{\mathbf{h}}_{k,n} - \bar{\mathbf{h}}_k)^H, 
\end{equation}
\begin{equation}
    \label{eq:covariance_matrix_final}
    \mathbf{R}_{\mathbf{e}_k} = \eta \mathbf{R}_{\mathbf{e}_k}^{S} + (1-\eta) \mathbf{R}_{\mathbf{e}_k}^{\text{Net}},
\end{equation}
where $\eta \in [0,1]$ is a weighting factor. This design strikes a balance, i.e., a larger $\eta$ favors high-quality empirical observations, while a smaller $\eta$ leverages the learned prior to counteract noise or sample scarcity. The resulting fused covariance matrices yield the second-order statistics in (\ref{eq:second_order_expectation}) and (\ref{eq:second_order_expectation_v}) for the robust WMMSE algorithm, thereby enhancing estimation reliability by synergizing data-driven prediction with empirical data. 

The specific training details of the covariance prediction network, along with the implementation of the MB-MAML strategy, will be elaborated in Section \ref{sec:MAML}. Besides, since the beamforming update equations in the online adaptation phase are identical to those derived above, we directly apply the same set of equations without repeating their derivation or notation hereafter.

\subsubsection{Online Adaptation Phase}
In the online phase, the network parameters $\boldsymbol{\theta}$ are fine-tuned to improve the subsequent covariance predictions $\mathbf{R}_{\mathbf{e}}^{\text{Net}}$. This adaptation is driven by a loss function directly linked to the system's performance metric, typically formulated as the negative average WSR and approximated via SAA, i.e.,
\begin{equation}
    \label{eq:loss_function}
    \mathcal{L}(\boldsymbol{\theta}) = -\sum_{n=1}^N \sum_{k=1}^K \omega_k
    \log_2 \left( 1 + \frac{|\hat{\mathbf{h}}_{k,n}^H \mathbf{v}_k|^2}{\sum_{i\neq k} |\hat{\mathbf{h}}_{k,n}^H \mathbf{v}_i|^2 + \sigma_k^2} \right).
\end{equation}
In particular, the parameters $\boldsymbol{\theta}$ are updated via gradient-based optimization methods such as Adam, where the gradient $\nabla_{\boldsymbol{\theta}} \mathcal{L}$ is computed through backpropagation and applied as
\begin{equation}
    \label{eq:online_update}
    \boldsymbol{\theta} \leftarrow \boldsymbol{\theta} - \alpha \nabla_{\boldsymbol{\theta}} \mathcal{L},
\end{equation}
where $\alpha$ denotes the online learning rate.
The effectiveness of this online fine-tuning mechanism stems from the inherent statistical differences between the offline learned model and the real-time observations. The parameters $\boldsymbol{\theta}$ inherited from the extensive offline training phase provide a covariance prediction that is highly stable but may suffer from bias due to the discrepancy between the current channel distribution and the offline training data. Conversely, the online objective function $\mathcal{L}(\boldsymbol{\theta})$ derived from instantaneous channel samples is inherently unbiased but exhibits high variance due to the limited sample size $N$. The online adaptation process effectively fuses these two perspectives: it leverages the stability of the offline knowledge while using the unbiased, high-variance online measurement to correct the systematic bias in the network's prediction. This crucial bias correction yields significant performance gains. Consequently, this online fine-tuning mechanism remains effective even when the number of instantaneous channel samples $N$ is small, e.g., $N=2$. The overall online robust beamforming procedure is summarized in Algorithm \ref{alg:online_robust_beamforming}.

\begin{algorithm}[t]
\caption{\small Online Robust Beamforming Design} \small
\label{alg:online_robust_beamforming}
\begin{algorithmic}[1]  
\STATE \textbf{Input:} Channel estimation samples $\{\hat{\mathbf{h}}_{k,n}\}$, neural network parameters $\boldsymbol{\theta}$, weighting factor $\eta$, maximum power $P_{\max}$, and priority weights $\{\omega_k\}$.
\STATE \textbf{Output:} Beamforming vectors $\{\mathbf{v}_k\}$.
\STATE Calculate the channel mean via $\bar{\mathbf{h}}_k = \frac{1}{N}\sum_{n=1}^{N}\hat{\mathbf{h}}_{k,n}$.
\STATE Compute the sample-based covariance matrix $\mathbf{R}_{\mathbf{e}_k}^S$ base on (\ref{eq:covariance_matrix_samples}).
\REPEAT
\STATE Predict the covariance matrix $\mathbf{R}_{\mathbf{e}_k}^{\text{Net}}$ using the neural network in (\ref{eq:covariance_matrix_network}).
\STATE Fuse the covariance matrices via (\ref{eq:covariance_matrix_final}) to obtain $\mathbf{R}_{\mathbf{e}_k}$.
\STATE Calculate the second-order statistics $\mathbb{E}\{\mathbf{h}_k \mathbf{h}_k^H\}$, $
\mathbb{E}\{\mathbf{h}_k^H \mathbf{v}_i \mathbf{v}_i^H \mathbf{h}_k\}$ using (\ref{eq:second_order_expectation}) and (\ref{eq:second_order_expectation_v}), respectively.
\STATE Update the beamforming $\mathbf{v}_k, \forall k$ by (\ref{eq:optimal_u_RW})-(\ref{eq:optimal_v_RW}) and normalize them to satisfy the power constraint $\sum_{k=1}^{K} \|\mathbf{v}_k\|^2 \le P_{\max}$.
\STATE Calculate the loss using (\ref{eq:loss_function}) and update the neural network parameters $\boldsymbol{\theta}$ by (\ref{eq:online_update}).
\UNTIL{\text{convergence}}
\end{algorithmic}
\end{algorithm} 

\subsection{Neural Network Design}
This subsection details the neural network design used in the covariance prediction module, including its architecture and the proposed SALR approximation technique to reduce computational complexity.

First, due to the inherent limitation of neural networks in processing complex-valued data, we decompose the input channel means $\{ \bar{\mathbf{h}}_k \}$ into their real and imaginary components as $[ \text{Re}\{ \bar{\mathbf{h}}_k \} , \text{Im}\{ \bar{\mathbf{h}}_k \}]$. 
Since the covariance matrix $\mathbf{R}_{\mathbf{e}_k}$ is constrained to be both real-valued and symmetric by definition, we can fully represent it by only learning its upper triangular elements; this allows us to define the learning function as $f_{\boldsymbol{\theta}}: \mathbb{R}^{2K M_t} \rightarrow  \mathbb{R}^{K  M_t (M_t+1)/2}$.
However, the output dimension of this network scales linearly with the number of users, presenting potential scalability challenges in large-scale deployment scenarios. To alleviate this issue, it is desirable to design a neural network on a per-user basis, where each user has its own neural network. This approach allows for independent processing of each user's channel estimation samples, enabling the network to learn user-specific characteristics and adapt to varying channel conditions. However, this design may lead to excessive computational overhead in scenarios with a large number of users. To address this issue, we adopt a shared neural network architecture that can process multiple users' data simultaneously while maintaining the ability to learn user-specific characteristics. By sharing the same neural network across users, we can significantly reduce the number of parameters from $KM_t(M_t+1)/2$ to $M_t(M_t+1)/2$. This shared network architecture allows us to efficiently process multiple users' data while maintaining a compact model size, making it suitable for large-scale deployment scenarios.
\subsubsection{Neural Network Architecture}
The shared network consists of a shallow DNN with three fully-connected layers and an output layer.
Due to the universal approximation theorem, a three-layer DNN is sufficient for function approximation. Although more advanced models, such as Transformers, could be used, this paper opts for a simpler three-layer DNN for convenience and as a representative example. This choice helps reduce the number of network parameters, thereby minimizing both training and inference computational overhead. Each fully-connected layer is followed by normalization and activation operations. The dimensions of the input layer and the two hidden layers are $D_1=2M_t$, $D_2=\lfloor M_t^2/4 \rfloor$ and $D_3= \lfloor M_t^2/2 \rfloor$ respectively. Besides, the dimension of the output layer is $ D_o = \lfloor M_t(M_t+1)/2 \rfloor$, which corresponds to the number of upper triangular elements of the covariance matrix $\mathbf{R}_{\mathbf{e}_k}$. The normalization layer employs the standard Batch Normalization technique, the activation function used in the hidden layers is ReLU, and the output layer uses a sigmoid activation function to ensure that the output can take any real value.

\subsubsection{Sparse Augmented Low-Rank Approximation}
To further achieve a favorable trade-off between model complexity and prediction accuracy, we introduce a structured approximation method via SALR approximation. The core idea is to represent $\mathbf{R}_{\mathbf{e_k}}$ as the sum of a low-rank component $ \sum_{i=1}^{r} \mathbf{a}_i \mathbf{a}_i^H$ and a sparse residual term $\mathbf{S}$, i.e.,
\begin{equation}
  \label{eq:salr_decomposition}
  \mathbf{R}_{\mathbf{e}_k} = \sum_{i=1}^{r} \mathbf{a}_i \mathbf{a}_i^H + \mathbf{S},
\end{equation}
where $r$ is the rank of the low-rank component, $\mathbf{a}_i \in \mathbb{C}^{M_t \times 1}$ are the basis vectors capturing the dominant correlations in the channel, and $\mathbf{S} \in \mathbb{C}^{M_t \times M_t}$ is a sparse symmetric matrix modeling weaker or localized variations with sparsity level $\delta = \frac{|\mathbf{S}|_0}{M_t^2}$, which is defined as the ratio of the number of non-zero entries $|\mathbf{S}|_0$ to the total number of entries in the matrix $M_t^2$. The sparsity pattern of $\mathbf{S}$ is determined offline by randomly generating several candidate masks and selecting the one that yields the best performance. Both $r$ and $\delta$ are chosen to be small (e.g., $r$=8 and $\delta$=0.09 are adopted in our simulations), enabling substantial reduction in the number of learnable parameters from $M_t(M_t+1/2)$ to $rM_t+\delta M_t^2$, especially when $\delta \ll 1$ and $r \ll 1$, while preserving sufficient expressive power. This hybrid representation allows the network to focus on the most important covariance structures, reduces overfitting risk, and lowers computational and memory requirements.
Note that a pure low-rank approximation, i.e., $\mathbf{S} = \mathbf{0}$, incurs non-negligible performance loss, particularly when $r$ is too small, as it fails to capture the residual channel variations. 
This trade-off is analogous to the one in low-rank adaptation (LoRA) for large language models, where efficiency gains come at the cost of expressivity.
A theoretical basis for our hybrid approach is provided in \cite{hanSLTrainSparseLowrank2024}, whose Proposition 1 proves that a low-rank matrix plus a uniformly sparse matrix with $\mathcal{O}(\log n / n)$ non-zero entries is full-rank with high probability.
This provides a principled foundation for using sparse compensation to recover the expressivity lost in low-rank approximations.
The values of $r$ and $\delta$ can be selected via cross-validation to balance complexity and performance for a given scenario. This enables real-time online adaptation while maintaining robustness under unknown and non-stationary channel statistics.
\begin{figure*}
\centering
\includegraphics[width=0.85\textwidth]{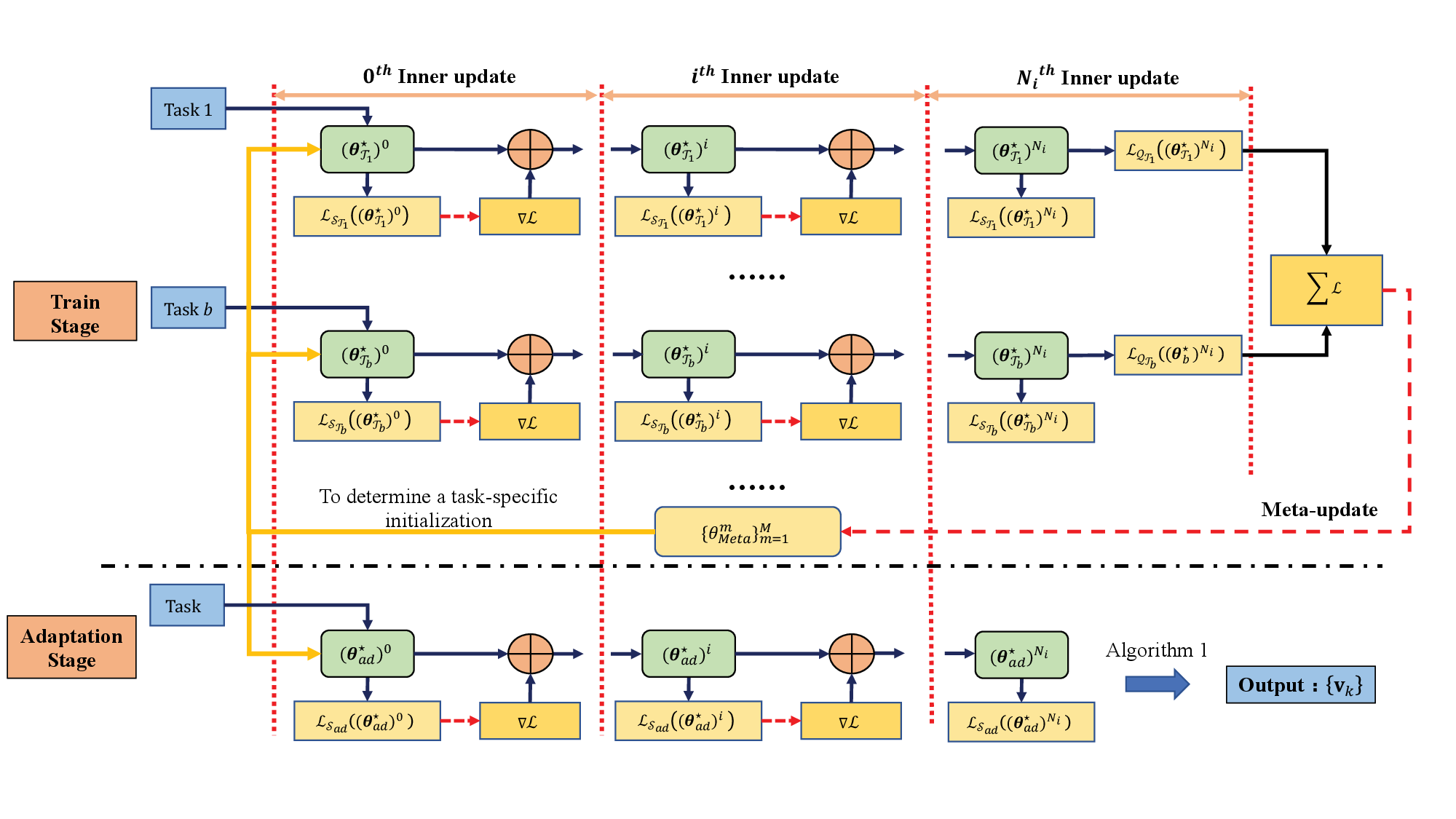} \vspace{-2em}
\caption{Proposed MB-MAML-based offline training and online adaptation scheme.}
\label{fig3_MB_MAML_framework}
\end{figure*}

\subsection{MB-MAML-Based Offline Learning and Online Adaptation}
\label{sec:MAML}
Despite enabling effective online adaptation, the proposed framework still faces two key limitations: 1) the fine-tuning process incurs a non-trivial computational overhead, as converging for each channel realization may require dozens to hundreds of gradient updates, and 2) conventional fine-tuning strategies often exhibit suboptimal generalization when encountering out-of-distribution channel error characteristics.
To overcome these limitations and inspired by the works \cite{MAML2017, leeXBMAMLLearningExpandable2024}, we extend the meta-learning paradigm to a multiple-basis paradigm, termed MB-MAML. Specifically, MB-MAML maintains a set of diverse initialized network parameters, also known as meta-bases, to enable efficient and robust adaptation across continuous and different channel conditions while substantially reducing online computation.

First of all, the proposed MB-MAML approach aligns the robust beamforming problem with the meta-learning paradigm by introducing the concept of a task.
Each task corresponds to a specific channel environment characterized by distinct channel estimation error statistics, thereby allowing the model to learn adaptation strategies that generalize across diverse channel conditions. Formally, a task $\mathcal{T}_b$ is defined using two sets drawn from the training data $\mathcal{D}_{\text{tr}}$. The support set $\mathcal{S}_{\mathcal{T}_b} = \left\{ \{\hat{\mathbf{h}}_{k,n}^{tr}\}_{n=1}^{N} \right\}_{k=1}^{K}$ of task $\mathcal{T}_b$ contains the channel estimation samples $\hat{\mathbf{h}}_{k,n}^{tr}$ used by the model for fine-tuning, and the query set $\mathcal{Q}_{\mathcal{T}_b} = \{ \mathbf{h}_{k}^{tr} \}_{k=1}^{K}$ of task $\mathcal{T}_b$ contains the corresponding true channel realizations $\mathbf{h}_{k}^{tr}$ for performance evaluation.
Based on these definitions, the MB-MAML framework operates through two main components in the offline training stage, including an inner loop and an outer loop, followed by an online fine-tuning stage.
During the offline training stage, the model is trained on a diverse set of meta-tasks and learns $M$ meta-bases $\{\boldsymbol{\theta}^{(m)}\}_{m=1}^{M}$ that minimize the expected adaptation loss across tasks, which will be introduced below. During the online fine-tuning stage, we first evaluate the performance of each meta-base ${\boldsymbol{\theta}^{(m)}}$ on the current task and then select the best-performing meta-base for fine-tuning. This fine-tuning is done using only a small number of gradient updates, typically 1 to 5 steps, achieving great performance at minimal computational cost. The overall MB-MAML framework is illustrated in Fig.~\ref{fig3_MB_MAML_framework} and the details are provided as follows.
\subsubsection{Offline Training Stage}
The objective of this stage is to learn the meta-bases $\{\boldsymbol{\theta}^{(m)}\}_{m=1}^{M}$ that enable rapid adaptation to unseen beamforming tasks and it is achieved by adopting a bi-level optimization structure: inner-loop adaptation and outer-loop meta-update. In each training iteration, we sample $B$ tasks to form a task set $\{\mathcal{T}_{b} \}_{b=1}^{B}$, inner-loop updates are first performed for every task in the set and outer-loop meta-update is then executed after all tasks have completed their inner-loop adaptations.
\paragraph{Inner-loop fine-tuning}
For each task $\mathcal{T}_{b}$, we execute the inner loop adaptation in three sequential steps.
First, to determine a task-specific initialization $\boldsymbol{\theta}_{\mathcal{T}_b}^\star$, we evaluate the suitability of all $M$ meta-bases $\{\boldsymbol{\theta}^{(m)}\}_{m=1}^{M}$ by computing the loss $\mathcal{L}_{\mathcal{S}_{\mathcal{T}_{b}}}({\boldsymbol{\theta}^{(m)}})$ of the $m$-th meta-basis on the support set $\mathcal{S}_{\mathcal{T}_{b}}$ of task $\mathcal{T}_{b}$, which is given by
\begin{multline}
    \label{eq:support_set_loss}
    \mathcal{L}_{\mathcal{S}_{\mathcal{T}_{b}}}({\boldsymbol{\theta}^{(m)}}) = \\ - \sum_{n=1}^N \sum_{k=1}^K \omega_k
    \log_2 \left( 1  + \frac{ |(\hat{\mathbf{h}}_{k,n}^{tr})^H\mathbf{v}_k(\boldsymbol{\theta}^{(m)})|^2}{\sum\limits_{i \neq k} |(\hat{\mathbf{h}}_{k,n}^{tr})^H\mathbf{v}_i(\boldsymbol{\theta}^{(m)})|^2 + \sigma_k^2} \right),
\end{multline}
where $\mathbf{v}_k(\boldsymbol{\theta}^{(m)})$ denotes the beamforming vector computed using Algorithm \ref{alg:online_robust_beamforming} with network parameters $\boldsymbol{\theta}^{(m)}$.

Second, based on these support-set losses $\{ \mathcal{L}_{\mathcal{S}_{\mathcal{T}_{b}}}({\boldsymbol{\theta}^{(m)}})\}$, we calculate some soft-selection weights $\sigma_{\mathcal{S}_{\mathcal{T}_b}}^{(m)}$ to combine the meta-bases $\{\boldsymbol{\theta}^{(m)}\}_{m=1}^{M}$, which are given as follows:
\begin{equation}
  \label{eq:soft_selection_weights_ieee}
\sigma_{\mathcal{S}_{\mathcal{T}_b}}^{(m)} = \frac{\exp(-\mathcal{L}_{\mathcal{S}_{\mathcal{T}_{b}}}(\boldsymbol{\theta}^{(m)}))}{\sum_{m'=1}^M \exp(-\mathcal{L}_{\mathcal{S}_{\mathcal{T}_{b}}}(\boldsymbol{\theta}^{(m')}))}.
\end{equation}
In particular, we employ the softmax function for this weighting mechanism to ensure critical optimization properties. In particular, the softmax function is fully differentiable, which is essential for end-to-end meta-learning, as its continuous and smooth mapping from loss to weight facilitates robust gradient propagation during the outer-loop meta-update, thereby enabling effective backpropagation through the interpolation mechanism. Furthermore, by taking the exponent of the negative loss as in (\ref{eq:soft_selection_weights_ieee}), the softmax operation ensures that lower loss values translate into exponentially higher weights. This high-contrast sensitivity dynamically emphasizes the most relevant meta-basis, enabling the optimization framework to quickly identify and leverage the best prior knowledge for the current task while maintaining a smooth optimization landscape. Using these weights, the task-specific initialization $\boldsymbol{\theta}_{\mathcal{T}_b}^\star$ for task $\mathcal{T}_b$ is obtained by interpolating the meta-bases $\{\boldsymbol{\theta}^{(m)}\}_{m=1}^{M}$ as 
\begin{equation}
  \label{eq:task_specific_initialization_ieee}
\boldsymbol{\theta}_{\mathcal{T}_b}^\star = \sum_{m=1}^M \sigma_{\mathcal{S}_{\mathcal{T}_b}}^{(m)} \boldsymbol{\theta}^{(m)}.
\end{equation}
This interpolation capability is crucial for enhancing generalization, i.e., it allows the model to dynamically compose a task-adaptive prior, ensuring accurate and fast adaptation even for new channel conditions ($\mathbf{R}_{\mathbf{e}}$ configurations) that were not explicitly represented in the finite training task set.

Third, starting from the interpolated task-specific initialization $\boldsymbol{\theta}_{\mathcal{T}_b}^\star$, the inner loop executes the critical phase of task-specific adaptation by applying a few gradient updates on the support set $\mathcal{S}_{\mathcal{T}_{b}}$. The support set loss $\mathcal{L}_{\mathcal{S}_{\mathcal{T}_{b}}}$ serves as the objective function because the inner loop's primary goal is to minimize the immediate prediction error on the available samples, thereby quickly aligning the meta-initialized network with the specific channel statistics of the current task. By performing only a small number of gradient steps (denoted by $N_i$), the network quickly refines its parameters to fit the unique characteristics of the current beamforming task, transforming the generalized meta-knowledge into task-specific expertise. The update rule is given by
\begin{equation}
  \label{eq:inner_loop_update}
(\boldsymbol{\theta}_{\mathcal{T}_b}^\star)^{i} = (\boldsymbol{\theta}_{\mathcal{T}_b}^\star)^{i-1} - \alpha \nabla_\theta \mathcal{L}_{\mathcal{S}_{\mathcal{T}_{b}}}((\boldsymbol{\theta}_{\mathcal{T}_b}^\star)^{i-1}), \quad i = 1,\dots,N_i.
\end{equation}
 A small $N_i$ the is key to the efficiency of the MB-MAML framework, ensuring that the heavy computational burden of convergence is primarily handled offline during the meta-training phase.

\paragraph{Outer-loop meta-update}
In the outer loop, the meta-bases $\{\boldsymbol{\theta}^{(m)}\}_{m=1}^{M}$ are updated based on the performance achieved after inner-loop adaptation, which is referred to as the meta-update. The meta-update represents a higher-level optimization task, aiming to find initialization parameters that yield the best performance on unseen query sets across diverse tasks. Specifically, the update rule for the $m$-th meta-basis is derived from an aggregated query loss and an orthogonality regularization term as follows:
\begin{align}
    \label{eq:outer_loop_update}
\boldsymbol{\theta}^{(m)} &= \boldsymbol{\theta}^{(m)} - \beta_{\text{Meta}} \nabla_{\boldsymbol{\theta}^{(m)}} \sum_{b=1}^B \sum_{i=1}^{N_i} \gamma_i \sigma_{\mathcal{T}_b}^{(m)} \big( \mathcal{L}_{\mathcal{Q}_{\mathcal{T}_b}}((\boldsymbol{\theta}_{\mathcal{T}_b}^{\star})^i) \notag \\
& \quad + \lambda_{reg} \mathcal{L}_{reg}^{(m)} \big),
\end{align}
where $\mathcal{L}_{\mathcal{Q}_{\mathcal{T}_b}}((\boldsymbol{\theta}_{\mathcal{T}_b}^{\star})^i)$ is the query loss defined as the negative WSR and given by
\begin{multline}
    \label{eq:query_set_loss}
    \mathcal{L}_{\mathcal{Q}_{\mathcal{T}_b}}((\boldsymbol{\theta}_{\mathcal{T}_b}^{\star})^i) = \\ - \sum_{k=1}^K \omega_k
    \log_2 \left( 1  + \frac{|({\mathbf{h}}_{k}^{tr})^H\mathbf{v}_k((\boldsymbol{\theta}_{\mathcal{T}_b}^{\star})^i)|^2}{\sum\limits_{i \neq k} |({\mathbf{h}}_k^{tr})^H\mathbf{v}_i((\boldsymbol{\theta}_{\mathcal{T}_b}^{\star})^i)|^2 + \sigma_k^2} \right),
\end{multline}
$\beta_{\text{Meta}}$ is the meta-learning rate, $\gamma_i$ is a step-wise weighting factor, $\lambda_{reg}$ is the regularization weight, and $\sigma_{\mathcal{S_{\mathcal{T}_b}}^{(m)}}$ is the soft-selection weight which ensures that the update for $\boldsymbol{\theta}^{(m)}$ is proportional to its contribution to the final performance.
Besides, $\mathcal{L}_{reg}^{(m)}$ is the orthogonality regularization term  that explicitly encourages functional diversity among the meta-bases by minimizing the inner product between distinct parameter vectors, which is given by
\begin{equation}
  \label{eq:regularization_loss}
\mathcal{L}_{reg}^{(m)} = \sum_{m' \neq m}^M \langle \boldsymbol{\theta}^{(m)}, \boldsymbol{\theta}^{(m')} \rangle.
\end{equation}
This regularization operation mitigates redundancy and promotes exploration of diverse initialization strategies, thereby enhancing the model's ability to adapt to a wide range of channel error statistics.

\subsubsection{Online Adaptation Stage}
Through the inner and outer loops of the offline training phase, we obtain \(M\) well-trained meta-bases $\{ (\boldsymbol{\theta}^{(m)}_{Meta})^i \}_{m=1}^{M}$. 
Upon encountering new channel conditions with unknown statistics during online adaptation, we propose to evaluate the support-set loss $\mathcal{L}_{\mathcal{S}_{ad}}(\boldsymbol{\theta}^{(m)}_{Meta})$ for each well-trained meta-basis $\boldsymbol{\theta}^{(m)}_{Meta}$ on the online adaptation set $\mathcal{S}_{ad} = \{ \{ \hat{\mathbf{h}}_{k,n}^{ad} \}_{n=1}^{N} \}_{k=1}^K$, which is defined as 
\begin{multline}
    \label{eq:online_support_set_loss}
    \mathcal{L}_{\mathcal{S}_{ad}}(\boldsymbol{\theta}^{(m)}_{Meta}) = \\ - \sum_{n=1}^{N} \sum_{k=1}^K \omega_k
    \log_2 \left( 1  + \frac{|(\hat{\mathbf{h}}_{k,n}^{ad})^H\mathbf{v}_k(\boldsymbol{\theta}^{(m)}_{Meta})|^2}{\sum\limits_{i \neq k} |(\hat{\mathbf{h}}_{k,n}^{ad})^H\mathbf{v}_i(\boldsymbol{\theta}^{(m)}_{Meta})|^2 + \sigma_k^2} \right).
\end{multline}
This evaluation directly guides the selection of the most promising initialization $\boldsymbol{\theta}_{ad}^\star$ for rapid fine-tuning, i.e., 
\begin{equation}
  \label{eq:optimal_initialization_selection}
\boldsymbol{\theta}_{ad}^\star = \arg \min_m \mathcal{L}_{\mathcal{S}_{ad}}(\boldsymbol{\theta}^{(m)}_{Meta}).
\end{equation}
Based on $\boldsymbol{\theta}_{ad}^\star$, a few gradient-based adaptation steps are further performed, which is given by 
\begin{equation}
  \label{eq:online_adaptation}
(\boldsymbol{\theta}_{ad}^{\star})^i = (\boldsymbol{\theta}_{ad}^{\star})^{i-1} - \alpha \nabla_{\boldsymbol{\theta}} \mathcal{L}_{\mathcal{S}_{ad}}((\boldsymbol{\theta}_{ad}^{\star})^{i-1}), \quad i = 1,\dots,N_i.
\end{equation}
The final adapted parameter $(\boldsymbol{\theta}_{ad}^{\star})^{N_i}$ is then used to generate the beamforming vectors for the current channel realization $\{ \mathbf{h}_k \}_{k=1}^K$. The combination of initialization selection and {rapid gradient-based adaptation} enables MB-MAML to generalize efficiently across the continuous and unbounded task space of $\mathbf{R}_{\mathbf{e}}$, without requiring retraining or statistical re-estimation, which is a key advantage over both conventional robust beamformers and vanilla meta-learning approaches. The complete hybrid offline-online robust beamforming design is summarized in Algorithm \ref{alg:maml_robust_beamforming}.

\begin{algorithm} 
\caption{\small Hybrid Offline-Online Robust Beamforming Design} \small
\label{alg:maml_robust_beamforming}   
\begin{algorithmic}[1]
\STATE Input: the training dataset $\mathcal{D}_{\text{tr}}$, the number of meta-bases $M$, the number of inner loop updates $N_i$, the online learning rate $\alpha$, the meta-learning rate $\beta_{Meta}$, and the regularization coefficient $\lambda_{reg}$.
\STATE Initialize: the initializations $\{ \boldsymbol{\theta}^{(m)}\}_{m=1}^M$. 

\textbf{Offline Training:}
\REPEAT
\STATE Sample a batch of $B$ tasks $\{\mathcal{T}_b \}_{b=1}^{B}$ from training dataset $\mathcal{D}_{\text{tr}}$.
\FOR{$b=1$ to $B$}    
\STATE Obtain support set $\mathcal{S}_{\mathcal{T}_{b}}$ and query set $\mathcal{Q}_{\mathcal{T}_{b}}$ from task $\mathcal{T}_{b}$.
\FOR{$m=1$ to $M$}
\STATE Compute the support loss $\mathcal{L}_{\mathcal{S}_{\mathcal{T}_{b}}}(\boldsymbol{\theta}^{(m)})$ using (\ref{eq:support_set_loss}) for the task $\mathcal{T}_{b}$ with basis $\boldsymbol{\theta}^{(m)}$.
\ENDFOR
\STATE Compute the softmax weights $\sigma_{\mathcal{S}_{\mathcal{T}_{b}}}^{(m)}$ using (\ref{eq:soft_selection_weights_ieee}).
\STATE Compute the final initialization parameter $\boldsymbol{\theta}_{\mathcal{T}_b}^{\star}$ using (\ref{eq:task_specific_initialization_ieee}).
\STATE initialize the network with $(\boldsymbol{\theta}_{\mathcal{T}_b}^{\star})^{(0)} = \boldsymbol{\theta}_{\mathcal{T}_b}^{\star}$.
\FOR{$i=1$ to $N_i$}
\STATE Perform inner loop update using (\ref{eq:inner_loop_update}).
\ENDFOR
\ENDFOR
\STATE Perform outer loop update for all bases using (\ref{eq:outer_loop_update}).
\UNTIL{convergence}
\STATE Output: the optimized bases $\{\boldsymbol{\theta}^{(m)}_{Meta}\}_{m=1}^M$.

\textbf{Online Adaptation:} 
\FOR{each new task $\mathcal{T}_{ad}$}
\STATE Sample support set $\mathcal{S}_{ad}$.
\STATE Select initialization $\boldsymbol{\theta}_{ad}^\star$ using (\ref{eq:optimal_initialization_selection}).
\STATE initial the network with $(\boldsymbol{\theta}_{ad}^{\star})^0= \boldsymbol{\theta}_{ad}^\star$.
\FOR{$i=1$ to $N_i$}
\STATE Perform online adaptation using (\ref{eq:online_adaptation}).
\ENDFOR
\STATE Using adapted network parameters $(\boldsymbol{\theta}_{ad}^{\star})^{N_i}$ to design the beamforming vectors $\{\mathbf{v}_k\}$ by Algorithm 1.
\STATE Output: the beamforming vectors $\{\mathbf{v}_k\}$.
\ENDFOR
\end{algorithmic}   
\end{algorithm}

\subsection{Computational Complexity Analysis}
Since offline meta-training is performed in a non-real-time setting, our complexity analysis focuses exclusively on the online adaptation stage, which must operate under strict latency constraints.
The computational complexity of the proposed framework is dominated by the following three components:
\begin{enumerate}
\item {Beamforming matrix and dual variable updates:} This includes the computation of the receive coefficients $\{u_k\}$, the weighting factors $\{w_k\}$, and beamforming vectors $\{ \mathbf{v}_k\}$ via the statistical WMMSE rules. The complexity of this component is $\mathcal{O}\left( N_i (M_t^3 + K M_t^2+ K M_t) \right)$.
\item {Forward propagation through the prediction neural network:} This involves evaluating the covariance estimation network to generate $\mathbf{R}_{\mathbf{e}_k}^{\text{Net}}$ from the input channel mean. Since the network consists of three fully connected layers with dimensions $D_1$, $D_2$, and $D_3$, and outputs a $D_o$-dimensional vectorized covariance representation, the complexity is $\mathcal{O}\left( N_i K (D_1 D_2 + D_2 D_3 + D_3 D_o) \right)$.
\item {Backward propagation for gradient computation:} This component involves calculating the gradients of the loss with respect to the network parameters $\boldsymbol{\theta}$ for online fine-tuning. Due to the symmetry of forward and backward passes in fully connected networks, its complexity is similar to that of forward propagation, which is $\mathcal{O}\left( N_i K (D_1 D_2 + D_2 D_3 + D_3 D_o) \right)$.
\end{enumerate}
Combining these components, the total complexity of the online adaptation stage is
\begin{equation*}
  \mathcal{O}\left( N_i \left[ M_t^3 + K M_t^2 + 2K (D_1 D_2 + D_2 D_3 + D_3 D_o) \right] \right).
\end{equation*}
The result shows that the complexity scales linearly with the number of adaptation steps $N_i$ and the number of users $K$, while remaining independent of the channel sample size, a key advantage over sample-averaging approaches such as SAA. The use of the SALR approach further reduces the orders of $D_1,D_2,D_3$ and $D_o$ to $O(M_t)$ by exploiting the inherent structure of the error covariance matrix, yielding additional computational savings in practice.
Consequently, the forward propagation complexity of the neural network is $\mathcal{O}(M_t^2)$ and the computational complexity for both the second and third components of the framework is $\mathcal{O}(K N_i M_t^2)$.

In contrast, the robust WMMSE algorithm, mentioned in Section \ref{sec:related work}, involves an iterative solution process where each iteration computes beamforming vectors. The complexity of robust WMMSE is given by $\mathcal{O}\left( N_i (M_t^3 + K M_t^2 + K M_t) \right)$,
which is similar to the complexity of our proposed framework. In comparison, the SWMMSE algorithm, which also computes beamforming vectors using the WMMSE algorithm in each iteration, has a complexity of $\mathcal{O}\left( N_i (M_t^3 + K N M_t^2 + K M_t) \right)$, 
which is higher due to the additional factor \(N\), corresponding to the number of channel estimation samples in the system. Furthermore, the number of iterations $N_i$ in SWMMSE is often quite large, leading to considerable computational overhead in practice.

\section{Simulation Results}
\label{sec:simulation_results}
In this section, we evaluate the performance of the proposed hybrid offline–online framework. The CSI is generated using the QuaDRiGa model \cite{jaeckelQuaDRiGa2014} under the 3GPP 3D UMa NLOS scenario at a center frequency of 6~GHz with 20~MHz bandwidth. The BS has $M_t=32$ antennas and serves $K=4$ single-antenna users randomly distributed within a cell of 200 m radius. Each time block is 0.5 ms in duration and comprises 10 slots.
All input data are generated using MATLAB and PyTorch 2.4.1 is used to implement the proposed algorithms. The number of channel samples is set to 2 and the weighting factor is set to 0.1. Table \ref{table_simulation_parameters} illustrates the parameters used in the simulation. 
To model a random and symmetric covariance matrix, the channel estimation covariance matrix is modeled as $\mathbf{R}_{\mathbf{e}_k} = \mathbf{Q} \boldsymbol{\Lambda} \mathbf{Q}^H$ where $\mathbf{Q}$ is a randn unitary matrix and the diagonal entries of $\boldsymbol{\Lambda}$ drawn from an exponential-Gaussian distribution. By regenerating $\boldsymbol{\Lambda}$ for each coherence block while keeping $\mathbf{Q}$ fixed, the channel error statistics evolve over time, thereby creating non-stationary error conditions across blocks. In our simulations, the channel error severity is quantified by channel estimation error level $\gamma = |\mathbf{h}_k| / |\mathbf{e}_k|$, which is defined as the ratio of the channel gain to the error magnitude, and SNR is defined as $\text{SNR}={P_{\max}}/{\sigma^2}$.

\begin{table}
\renewcommand{\arraystretch}{1.3}
\caption{Simulation Parameters}
\label{table_simulation_parameters}
\centering
\begin{tabular}{|c|c|}

\hline
\textbf{Parameter} & \textbf{Value} \\
\hline
Number of channel samples ($N$) & 2 \\
\hline
Weighting factor ($\eta$) & 0.1 \\
\hline
Number of iterations ($N_i$) & 5 \\   
\hline
Number of epochs & 20 \\
\hline
dimension of hidden layer 1 ($D_1$) & 128 \\
\hline
dimension of hidden layer 2,3 ($D_2, D_3$) & 256 \\
\hline
dimension of output layer ($D_o$) & $rM_t+\delta M_T^2$ \\
\hline
Meta learning rate ($\beta_{Meta}$) & 0.001 \\
\hline
Online learning rate ($\alpha$) & 0.01 \\
\hline
Batch size ($N_b$) & 20 \\
\hline
Number of tasks in each batch ($B$) & 20 \\
\hline
Low rank factor ($r$) & 8 \\
\hline
Sparsity factor ($\delta$) & 0.09 \\
\hline
Number of initializations basis ($M$) & 8 \\
\hline
Regularization coefficient ($\lambda_{reg}$) & 0.001 \\
\hline
\end{tabular}
\end{table}

\subsection{Convergence Behavior}
\begin{figure}
  \centering
  \includegraphics[width=0.4\textwidth]{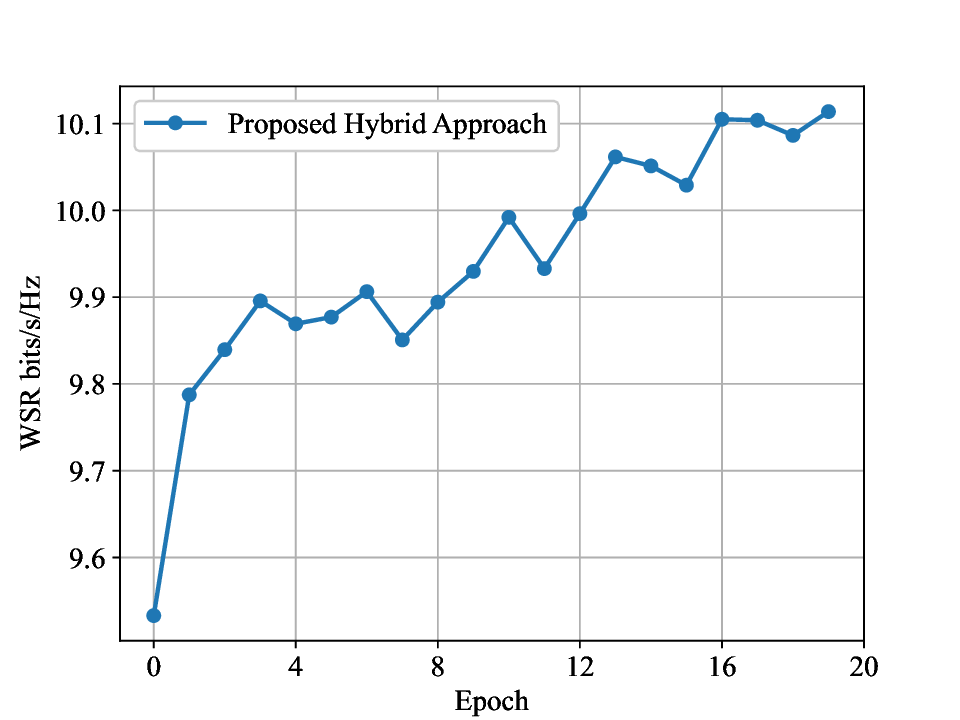} 
  \caption{Convergence behavior of the proposed hybrid approach.}  
  \label{fig:Convergence_behavior}
\end{figure}

Fig.~\ref{fig:Convergence_behavior} illustrates the convergence behavior of the proposed hybrid offline-online framework. The WSR performance in the figure is obtained by taking the network parameter $\{ \boldsymbol{\theta}^{(m)} \}_{m=1}^M$ learned at each epoch, performing $N_i=5$ online adaptation steps, computing the beamforming vectors based on the adapted parameters $\{ \boldsymbol{\theta}^{\star}_{ad} \}^{N_i}$, and evaluating the resulting WSR. As training progresses, the WSR achieved by the resulting beamforming vectors steadily improves and eventually converges to a steady performance level, reflecting the stable convergence behavior of the proposed framework.

\begin{figure}
  \centering
  \includegraphics[width=0.4\textwidth]{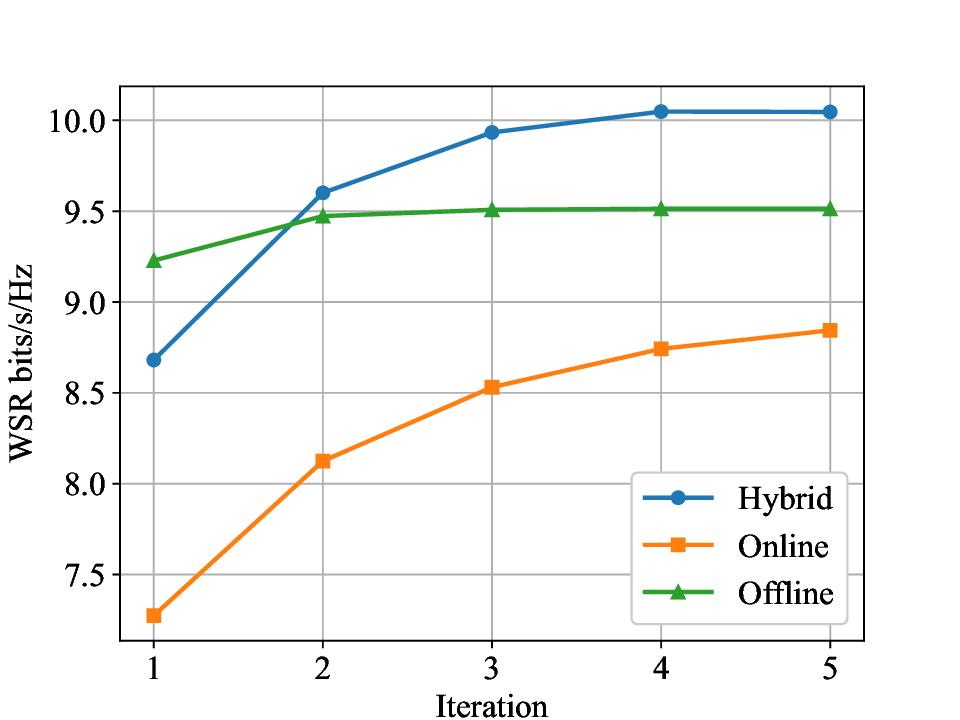}  
  \caption{Performance Comparison of Offline, Online, and Hybrid approach.}
  \label{fig:Effect of Offline-Online Hybrid}
\end{figure}

\begin{figure*}
    \centering
        \subfloat[Impact of Sparsity Factor $\delta$]{
        \includegraphics[width=0.4\textwidth]{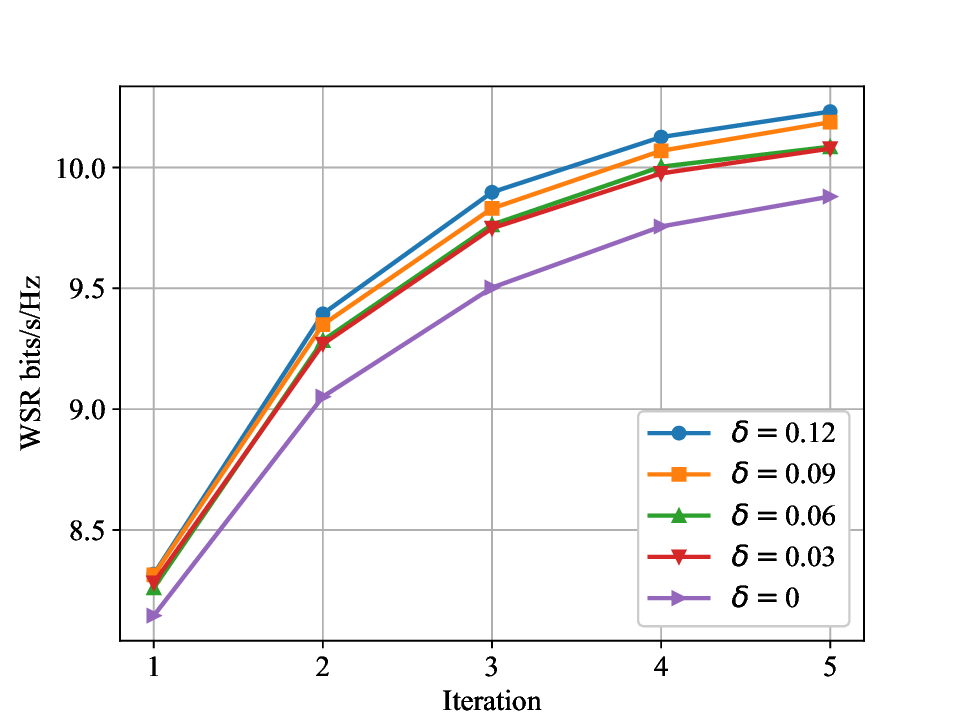}
        \label{fig:Effect_of_SALR_sparsity}
    }
    \hfil
    \subfloat[Performance-Complexity Trade-off of SALR]
    { \includegraphics[width=0.4\textwidth]{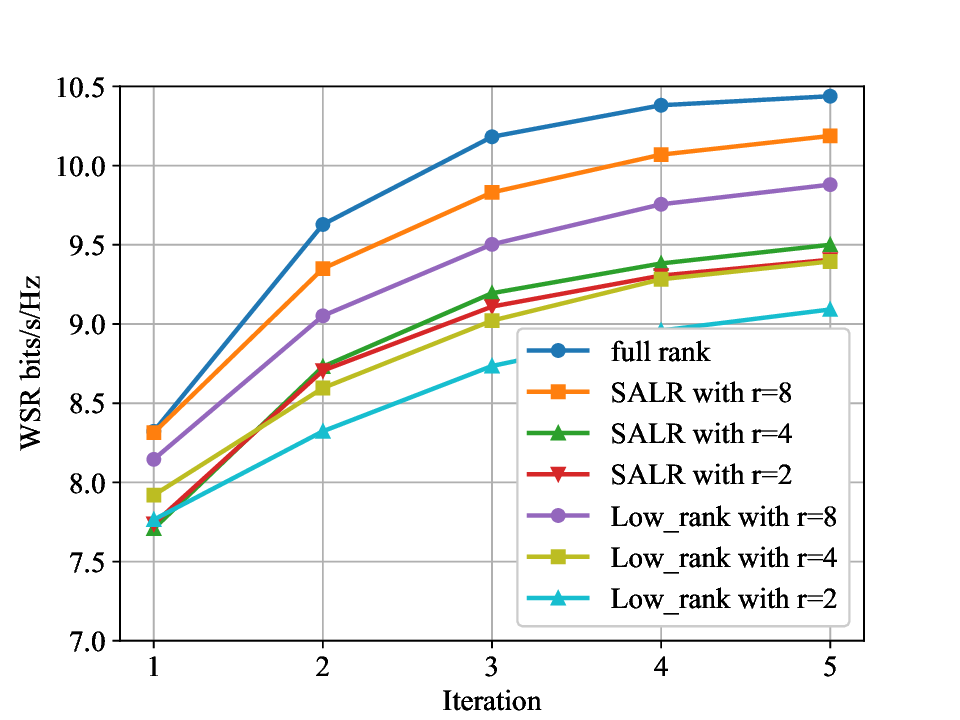}
        \label{fig:Effect_of_SALR_rank}
    }
    \hfil
      \caption{WSR performance comparison under different ranks and sparsity levels.}
  \label{fig:WSR_SALR_comparison}
\end{figure*}

\begin{figure*}
\centering \vspace{-1em}
 \subfloat[Effect of the Number of Meta-Initializations]{\includegraphics[width=0.4\textwidth]{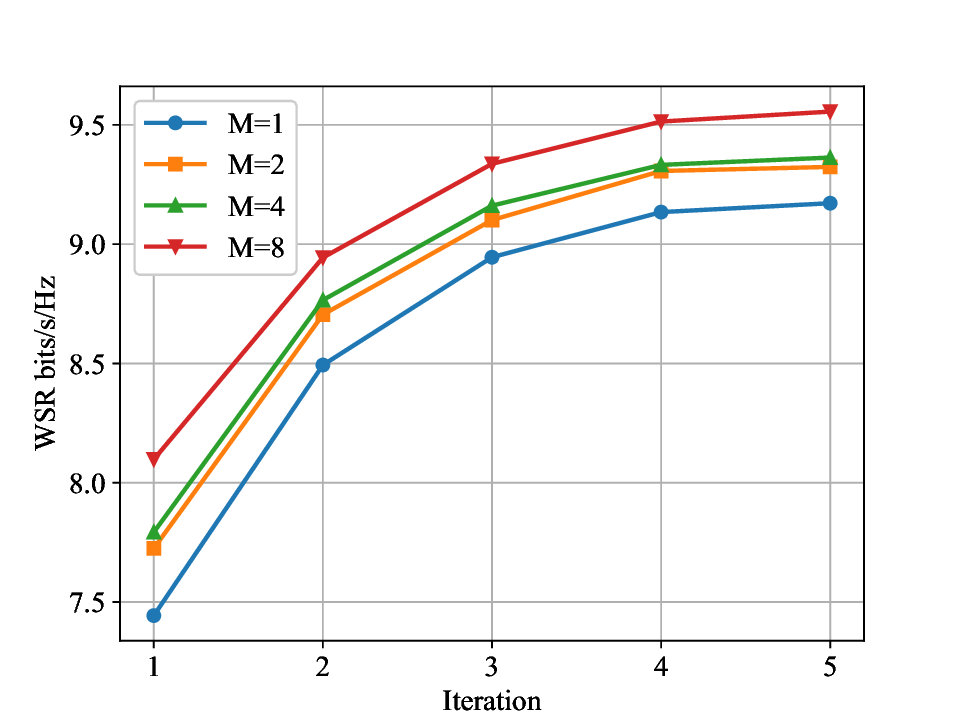} \label{fig:Effect of multiple basis}}
\hfil
 \subfloat[ID and OOD Generalization Comparison]{\includegraphics[width=0.4\textwidth]{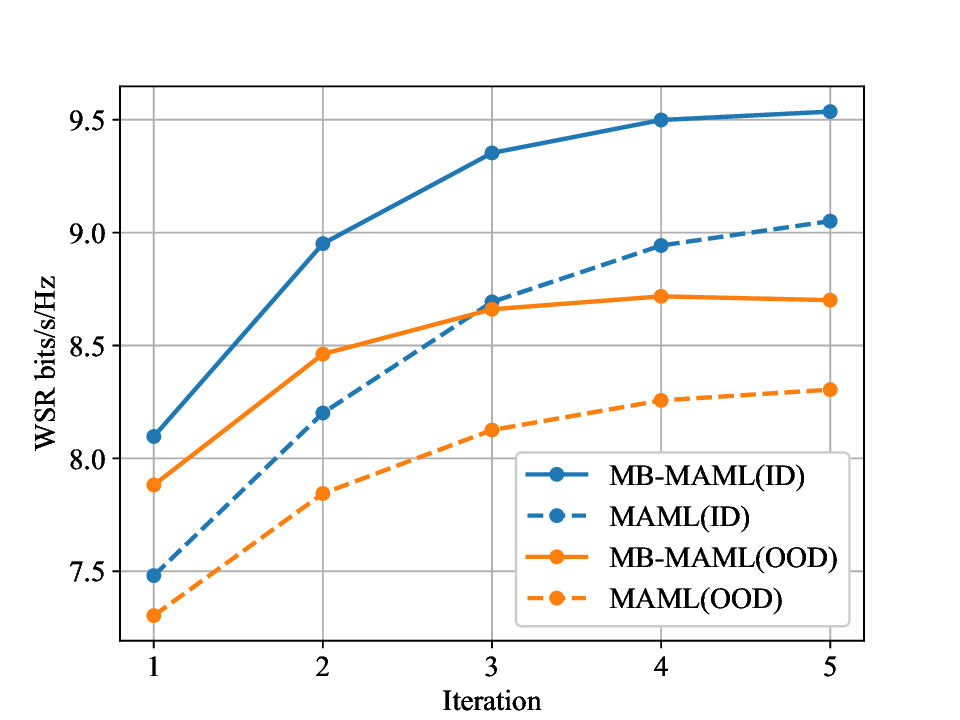} \label{fig:Test of ODD}}
  \hfil
    \caption{Impact of multiple meta-initializations on WSR performance and generalization.}
\label{fig:design_performance}
\end{figure*}
\subsection{Ablation Studies}
\label{subsec:ablation}
We begin with an ablation study to isolate and evaluate the individual contributions of the components in the proposed hybrid offline–online approach.
Fig.~\ref{fig:Effect of Offline-Online Hybrid} compares the proposed hybrid offline-online approach with two ablated variants: (i) a purely offline approach, where the network parameters $\boldsymbol{\theta}$ are learned entirely through offline training and remain fixed during deployment without any online fine-tuning, and (ii) a purely online approach, which performs online fine-tuning starting from randomly initialized parameters using only online gradient updates. The results demonstrate that although the purely online approach shows rapid initial gains in the first few steps, its performance ultimately fails to reach a satisfactory level due to poor initialization.
The purely offline approach, on the other hand, achieves high initial performance but shows limited improvement during online deployment, as it lacks the capability for task-specific adaptation. In contrast, the proposed hybrid framework effectively combines offline meta-training with online fine-tuning, and it is able to achieve superior WSR performance across all adaptation steps, validating that neither offline initialization nor online adaptation alone can achieve good performance.

Fig.~\ref{fig:WSR_SALR_comparison} illustrates the WSR performance achieved by employing the proposed SALR method under different ranks and sparsity levels. In particular, Fig.~\ref{fig:WSR_SALR_comparison} \subref{fig:Effect_of_SALR_sparsity} shows that the WSR improves as the sparsity factor $\delta$ increases when the low rank factor $r$ is set to 8; however, the growth rate diminishes as $\delta$ increases, eventually reaching a plateau. This indicates that even a very small sparsity factor is sufficient to yield noticeable performance gains. By balancing performance improvement and computational complexity, we set the sparsity factor to $\delta = 0.09$ in our implementations.
Fig.~\ref{fig:WSR_SALR_comparison} \subref{fig:Effect_of_SALR_rank} assesses the trade-off between approximation performance and computational efficiency of the proposed SALR method. The following two variants are compared: (i) full-rank estimation, (ii) pure low-rank approximation, i.e., $\mathbf{S}=0$. It is observed that, the WSR increases monotonically with the rank of the low-rank component, at the cost of higher computational complexity. While the full-rank approach achieves the best performance, its complexity is prohibitive for real-time deployment. The pure low-rank approximation approach offers significant complexity reduction but incurs notable performance loss. In contrast, the proposed SALR method, which augments the low-rank component with a learnable sparse residual, achieves substantially higher WSR than the pure low-rank approximation method under comparable complexity. This demonstrates that sparse compensation effectively recovers the expressivity lost in low-rank modeling, enabling an improved performance-complexity trade-off.

Fig.~\ref{fig:design_performance} \subref{fig:Effect of multiple basis} evaluates the impact of the number of meta-initializations $M$ on WSR performance and, more importantly, on generalization capability, where the low rank factor $r$ is set to 2 and the sparsity factor $\delta$ is set to 0. Under in-distribution (ID) testing, where the covariance matrices $\mathbf{R}_{\mathbf{e}_k}$ in the online adaptation set $\mathcal{S}_{ad}$ share the same eigenvectors $\mathbf{Q}$ as those in the training set $\mathcal{S}_{tr}$ and their eigenvalue matrices $\boldsymbol{\Lambda}$ are drawn from the same distribution, the results show that increasing $M$ consistently improves performance. This improvement arises because a larger number of meta-initializations provides broader task coverage within the training manifold.
In Fig.~\ref{fig:design_performance} \subref{fig:Test of ODD}, the proposed MB-MAML approach is compared against its single-initialization MAML counterpart under both ID and out-of-distribution (OOD) channel conditions where the number of meta-initializations $M$ is set to 8. As shown, under OOD testing, i.e., when faced with unseen channels with different spatial correlations or error strengths, the MB-MAML approach significantly reduces its generalization gap. This confirms that maintaining multiple diverse meta-initializations allows the proposed framework to adapt to a broader range of channel statistics, thereby improving robustness to distribution shifts, a critical requirement in practical deployments where channel conditions are often non-stationary or unknown. The ablation studies confirm that each component, hybrid adaptation, SALR approximation, and multiple initializations, contributes uniquely to performance.

\begin{figure*}
	\centering
	\subfloat[Sum rate versus SNR]{\includegraphics[width=0.45\textwidth]{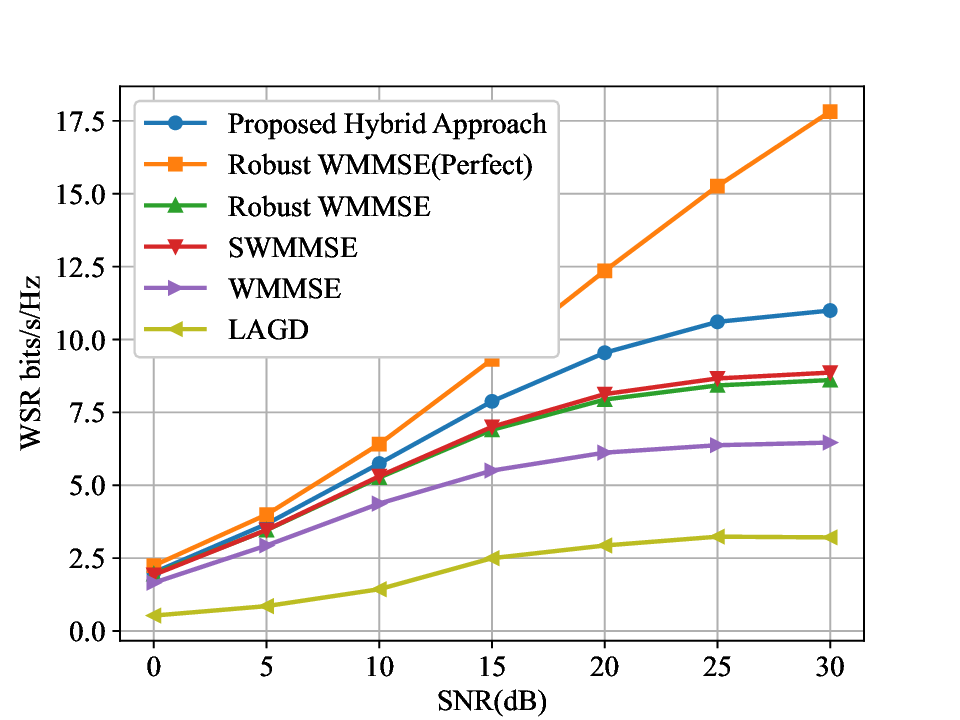} \label{fig:sum_rate_SNR}} 
	\hfil
	\subfloat[Sum rate versus channel estimation error]{\includegraphics[width=0.45\textwidth]{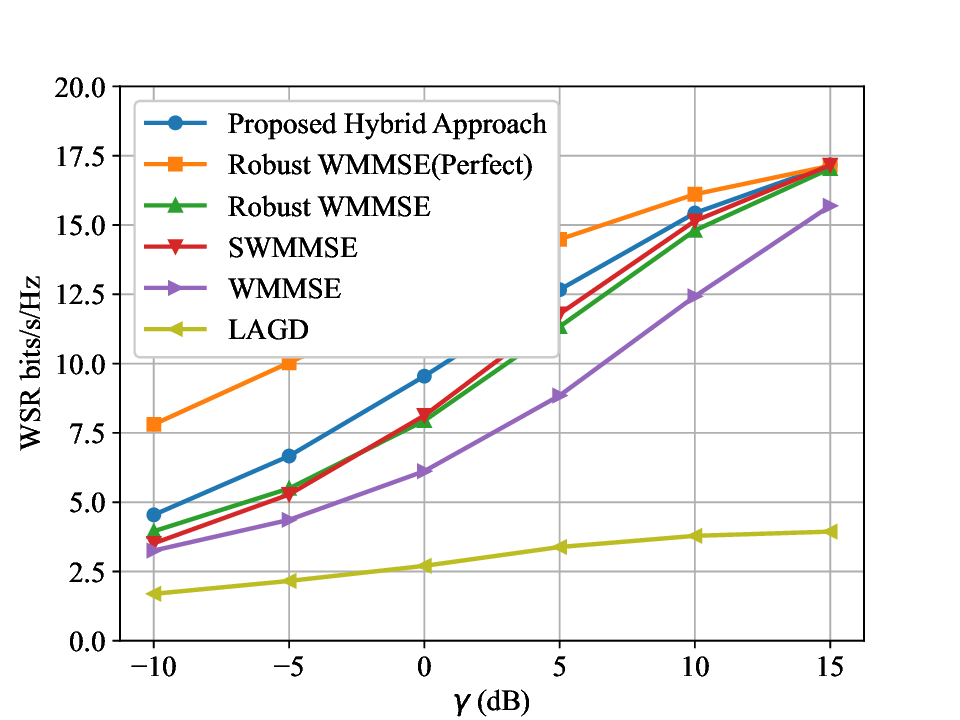} \label{fig:sum_rate_channel_estimation_error}}
	\caption{Performance comparison of the proposed hybrid offline-online approach with benchmark methods.}
	\label{fig:performance_comparison}
\end{figure*}

\subsection{Performance Comparison with State-of-the-Art Methods}
\label{subsec:sota_comparison}
Next, we evaluate the WSR performance of the proposed framework against four representative baseline methods:
\begin{itemize}
    \item \textbf{WMMSE} \cite{shi2011iteratively}: The classical non-robust beamforming method that assumes perfect CSI. Since perfect CSI is not available in practice, the empirical channel mean $\bar{\mathbf{h}}_k$ is used as the input.

    \item \textbf{SWMMSE}: The SAA-based robust beamforming algorithm introduced in Section.~\ref{sec:related work}, which requires a set of channel samples $\{ \mathbf{\hat{h}}_{k,n}\}_{n=1}^N$ for stochastic optimization. 

    \item \textbf{Robust WMMSE} \cite{shiRobustWMMSEPrecoder2023}: A statistically robust beamforming approach that assumes knowledge of the channel estimation error covariance matrix $\mathbf{R}_{\mathbf{e}}$. This method utilizes both the empirical channel mean $\bar{\mathbf{h}}_k$ and the estimated error covariance matrix $\mathbf{R}_{\mathbf{e}_k}^S$ as input features, where 
    
    \item \textbf{LAGD} \cite{LAGD2022}: A learning-assisted gradient descent (LAGD) approach that uses only the empirical channel mean $\bar{\mathbf{h}}_k$ as input.
\end{itemize}
To establish an oracle performance bound, the robust WMMSE algorithm utilizing the true channel error covariance matrix $\mathbf{R}_{\mathbf{e}_k}$ is also included in the simulations and is denoted as robust WMMSE (perfect). In our simulations, the channel mean and error covariance matrix for these methods are estimated using the sample mean and sample covariance based on the available channel samples $\{ \hat{\mathbf{h}}_{k,n} \}_{n=1}^N$. The number of channel samples $N$ is set to $N=2$ unless otherwise specified.

Fig.~\ref{fig:performance_comparison} \subref{fig:sum_rate_SNR} illustrates the WSR performance achieved by the considered approaches versus SNR where $\gamma=0$ dB. As shown, the proposed approach consistently achieves higher WSR than the consider baselines across the entire SNR range. At the low SNR regime, all methods exhibit comparable performance as the system is noise-dominated. However, as SNR increases, effective interference management becomes critical. Here, the superior interference mitigation capabilities of our hybrid offline-online design, powered by robust meta-learning initialization, become clearly evident. In contrast, both standard WMMSE and the LAGD approach exhibit substantial performance degradation at high SNR, which underscores the necessity of explicitly incorporating channel estimation errors into the beamforming design.

Fig.~\ref{fig:performance_comparison} \subref{fig:sum_rate_channel_estimation_error} shows the impact of channel estimation error level $\gamma$ on the WSR performance where SNR=20dB. It is observed that the proposed framework demonstrates superior WSR performance across various error levels, which confirms that it can effectively capitalize on limited channel samples to refine both covariance estimation and beamforming. Besides, at very high $\gamma$, all methods converge toward the oracle as the channel estimation errors become negligible, whereas under severe uncertainty (very low $\gamma$), their performance degrades noticeably. Nonetheless, the proposed approach maintains a consistent advantage, which shows that the hybrid offline–online framework achieves an effective balance between adaptability and robustness, outperforming existing state-of-the-art (SOTA) methods under realistic channel conditions.
The comparison with SOTA methods demonstrates that the proposed framework achieves superior WSR, especially under imperfect CSI and unknown statistics, while maintaining low online complexity. Crucially, the MB-MAML design ensures strong generalization to unseen channel conditions, making it a practical solution for real-world deployments.

\section{Conclusion}
\label{sec:conclusion}
This paper presented a hybrid offline–online framework for robust beamforming in MU-MIMO systems under unknown channel statistics. Unlike conventional methods that rely on analytical priors, the proposed approach achieves prior-free robustness by combining offline meta-learning with lightweight online adaptation.
In the offline phase, a shared DNN is trained to infer channel error covariance matrices directly from observed samples, enabling robustness across diverse channel conditions. To further reduce complexity, the SALR approximation provides an efficient representation of the predicted covariance matrices with minimal performance loss.
In the online phase, the network parameters are fine-tuned using current channel observations, allowing rapid adaptation to new channel distributions. Adaptation efficiency is enhanced through proposed MB-MAML strategy that supplies multiple meta-initializations as favorable starting points for online updates.
Extensive simulations verify that the hybrid framework, together with SALR and MB-MAML, improves WSR performance while reducing computational complexity and enabling fast adaptation. The proposed framework consistently outperforms SOTA baselines, demonstrating strong robustness and generalization under practical CSI uncertainty.
\bibliographystyle{IEEEtran}
\bibliography{IEEEabrv,ref_abrv}
\end{document}